\documentclass[acmsmall]{acmart}

\AtBeginDocument{%
  }

\usepackage{tikz}
\usetikzlibrary{mindmap}
\usepackage{metalogo}
\usepackage{smartdiagram}
\usesmartdiagramlibrary{additions}
\usepackage{forest}
\usetikzlibrary{shadows}
\usepackage{cleveref}
\usepackage{booktabs}
\usepackage{multirow}
\usepackage{makecell}
\usepackage{subfigure}
\usepackage[misc]{ifsym} 
\usepackage{balance}
\usepackage{listings}
\usepackage{color, xcolor}
\usepackage[linesnumbered,ruled,vlined]{algorithm2e}
\usepackage{tcolorbox}
\usepackage{amsmath}
\usepackage{mathtools}

\definecolor{lightcoral}{rgb}{0.94, 0.5, 0.5}
\definecolor{lightgreen}{rgb}{0.56, 0.93, 0.56}
\definecolor{harvestgold}{rgb}{0.85, 0.57, 0.0}
\definecolor{brightlavender}{rgb}{0.75, 0.58, 0.89}
\definecolor{capri}{rgb}{0.0, 0.75, 1.0}
\definecolor{carminepink}{rgb}{0.92, 0.3, 0.26}
\definecolor{celadon}{rgb}{0.67, 0.88, 0.69}
\definecolor{darkpastelgreen}{rgb}{0.01, 0.75, 0.24}
\definecolor{DeepSkyBlue4}{RGB}{0,104,139}

\setcopyright{acmlicensed}
\copyrightyear{2025}
\acmYear{2025}
\acmDOI{XXXXXXX.XXXXXXX}

\acmJournal{JACM}
\acmVolume{37}
\acmNumber{4}
\acmArticle{111}
\acmMonth{8}

\begin{document}

\title{A Survey of AIOps in the Era of Large Language Models}

\author{Lingzhe Zhang}
\affiliation{%
	\institution{Peking University}
	\city{Beijing}
	\country{China}}
\email{zhang.lingzhe@stu.pku.edu.cn}

\author{Tong Jia$^{\ast}$}
\thanks{*Corresponding author}
\affiliation{%
	\institution{Peking University}
	\city{Beijing}
	\country{China}}
\email{jia.tong@pku.edu.cn}

\author{Mengxi Jia}
\affiliation{%
	\institution{Peking University}
	\city{Beijing}
	\country{China}}
\email{mxjia@pku.edu.cn}

\author{Yifan Wu}
\affiliation{%
	\institution{Peking University}
	\city{Beijing}
	\country{China}}
\email{yifanwu@pku.edu.cn}

\author{Aiwei Liu}
\affiliation{%
	\institution{Tsinghua University}
	\city{Beijing}
	\country{China}}
\email{liuaw20@mails.tsinghua.edu.cn}

\author{Yong Yang}
\affiliation{%
	\institution{Peking University}
	\city{Beijing}
	\country{China}}
\email{yang.yong@pku.edu.cn}

\author{Zhonghai Wu}
\affiliation{%
	\institution{Peking University}
	\city{Beijing}
	\country{China}}
\email{wuzh@pku.edu.cn}

\author{Xuming Hu}
\affiliation{
	\institution{The Hong Kong University of Science and Technology (Guangzhou)}
	\city{Guangzhou}
	\country{China}}
\email{xuminghu@hkust-gz.edu.cn}

\author{Philip S. Yu}
\affiliation{
	\institution{University of Illinois Chicago}
	\city{Chicago}
	\country{United States}}
\email{psyu@uic.edu}

\author{Ying Li$^{\ast}$}
\affiliation{%
	\institution{Peking University}
	\city{Beijing}
	\country{China}}
\email{li.ying@pku.edu.cn}

\renewcommand{\shortauthors}{Lingzhe Zhang et al.}

\begin{abstract}
As large language models (LLMs) grow increasingly sophisticated and pervasive, their application to various Artificial Intelligence for IT Operations (AIOps) tasks has garnered significant attention. However, a comprehensive understanding of the impact, potential, and limitations of LLMs in AIOps remains in its infancy. To address this gap, we conducted a detailed survey of LLM4AIOps, focusing on how LLMs can optimize processes and improve outcomes in this domain. We analyzed 183 research papers published between January 2020 and December 2024 to answer four key research questions (RQs). In RQ1, we examine the diverse failure data sources utilized, including advanced LLM-based processing techniques for legacy data and the incorporation of new data sources enabled by LLMs. RQ2 explores the evolution of AIOps tasks, highlighting the emergence of novel tasks and the publication trends across these tasks. RQ3 investigates the various LLM-based methods applied to address AIOps challenges. Finally, RQ4 reviews evaluation methodologies tailored to assess LLM-integrated AIOps approaches. Based on our findings, we discuss the state-of-the-art advancements and trends, identify gaps in existing research, and propose promising directions for future exploration.
\end{abstract}




\maketitle

\section{Introduction}

Modern software systems are growing increasingly complex and typically serve massive user bases numbering in the billions~\cite{kang2022separation, zhang2024time}. Even minor software glitches can lead to substantial financial and reputational losses due to service interruptions or degraded performance~\cite{elliot2014devops}. As a result, large-scale distributed systems must ensure continuous 24/7 operation, with high demands for availability and reliability. However, the sheer scale and intricate logic of these systems make failures both inevitable and difficult to detect, localize, and diagnose. Once faults occur, analysis and recovery become even more challenging. Consequently, effective AIOps—encompassing rapid failure detection, efficient diagnosis, accurate root cause identification, and timely remediation—has become essential for maintaining system reliability and availability.

\noindent\textbf{Scope of AIOps in this paper.} Artificial Intelligence for IT Operations (AIOps) was first introduced by Gartner in 2016~\cite{prasad2018market}. AIOps refers to the use of machine learning (ML) and deep learning (DL) techniques to process vast amounts of data generated by operational tools and infrastructure, enabling the real-time detection, diagnosis, and resolution of system issues. This greatly enhances the automation and intelligence of IT operations. Most existing literature adopts this scope of AIOps, focusing on its application to operational reliability and failure management~\cite{notaro2021survey, cheng2023ai, remil2024aiops, wei2024log, su2024large}. However, some works have extended the definition of AIOps to include topics such as Ops for AI and network security~\cite{diaz2023joint, nguyen2024network}. These directions, while valuable, are beyond the scope of this paper. We focus specifically on the role of AIOps in addressing runtime software failures in large-scale distributed systems.

\subsection{Why are LLMs Beneficial for AIOps?}
\label{sec: llm4fm}

With the rapid advancement of Artificial Intelligence (AI), traditional AIOps approaches based on machine learning (ML) and deep learning (DL) have played a pivotal role in automated software failure management~\cite{notaro2021survey, diaz2023joint, cheng2023ai, remil2024aiops, wei2024log}. However, in practical applications, these methods still face several challenges:

\begin{itemize}
	\item \textbf{Need for complex feature extraction engineering.} These methods typically require extensive preprocessing and feature engineering to extract useful information from raw data. They have limited capabilities in understanding and processing data, especially in handling unstructured data such as logs and traces, which appearing relatively weak.
	\item \textbf{Lack of cross-platform generality.} Traditional models are often tuned and trained specifically for a particular software system. Once a different software system is adopted or even minor changes are made to the original system, the performance of the model significantly deteriorates, even when performing the same task.
	\item \textbf{Lack of cross-task flexibility.} Due to the singularity of model knowledge and outputs, models can only perform one task at a time. For example, in Root Cause Analysis (RCA) tasks, some work is aimed at identifying the cause of the problem~\cite{sui2023logkg, yuan2019approach, lin2016log}, while others are focused on identifying the software components involved~\cite{misiakos2024learning, ikram2022root, wang2023root}. In real-world scenarios, multiple models must run simultaneously to complete the entire RCA task.
	\item \textbf{Limited model adaptability.} With changes in the software system, deep learning-based methods typically require frequent model training and updates to adapt to new data and environments. While there are many online learning methods~\cite{ahmed2021anomaly, lyu2023assessing, li2023few, han2021log} available to address this issue, this process not only consumes time and effort but also can result in delayed responses from the model when handling sudden events.
	\item \textbf{Restricted levels of automation.} Current deep learning-based methods exhibit relatively limited capabilities in terms of automated operations and intelligent decision-making. While some degree of automation is achievable, significant manual intervention and configuration are still required. Particularly in the case of Auto Remediation, current efforts are mainly stopped at Incident Triage~\cite{zeng2017knowledge} or Solution Recommendation~\cite{ wang2017constructing, zhou2016resolution}.
\end{itemize}

Large Language Models (LLMs), pre-trained on natural language understanding tasks, offer a promising avenue for addressing these limitations. (1) Due to their robust natural language processing capabilities, LLMs can efficiently handle and comprehend unstructured data, often eliminating the need for prior feature extraction. (2) Trained on vast amounts of cross-platform data, LLMs possess a strong degree of generality and logical reasoning abilities. (3) Outputting natural language, LLMs offer great flexibility, enabling them to simultaneously perform multiple AIOps tasks, such as identifying the cause of a problem and the involved software components. (4) Leveraging their pre-training, LLMs exhibit powerful adaptive capabilities and can incorporate continuously updated external knowledge using methods like Retrieval-Augmented Generation (RAG), often without requiring retraining. (5) With robust script generation capabilities and the ability to automatically invoke external tools, LLMs can achieve higher levels of automation.

\subsection{Why a Survey of AIOps in the Era of LLMs? }

\begin{figure}[htbp]
	\centering
	\subfigure[Number of LLM-based publications in the field of AIOps]{
		\begin{minipage}{0.46\linewidth}
			\centering   
			\includegraphics[width=\textwidth]{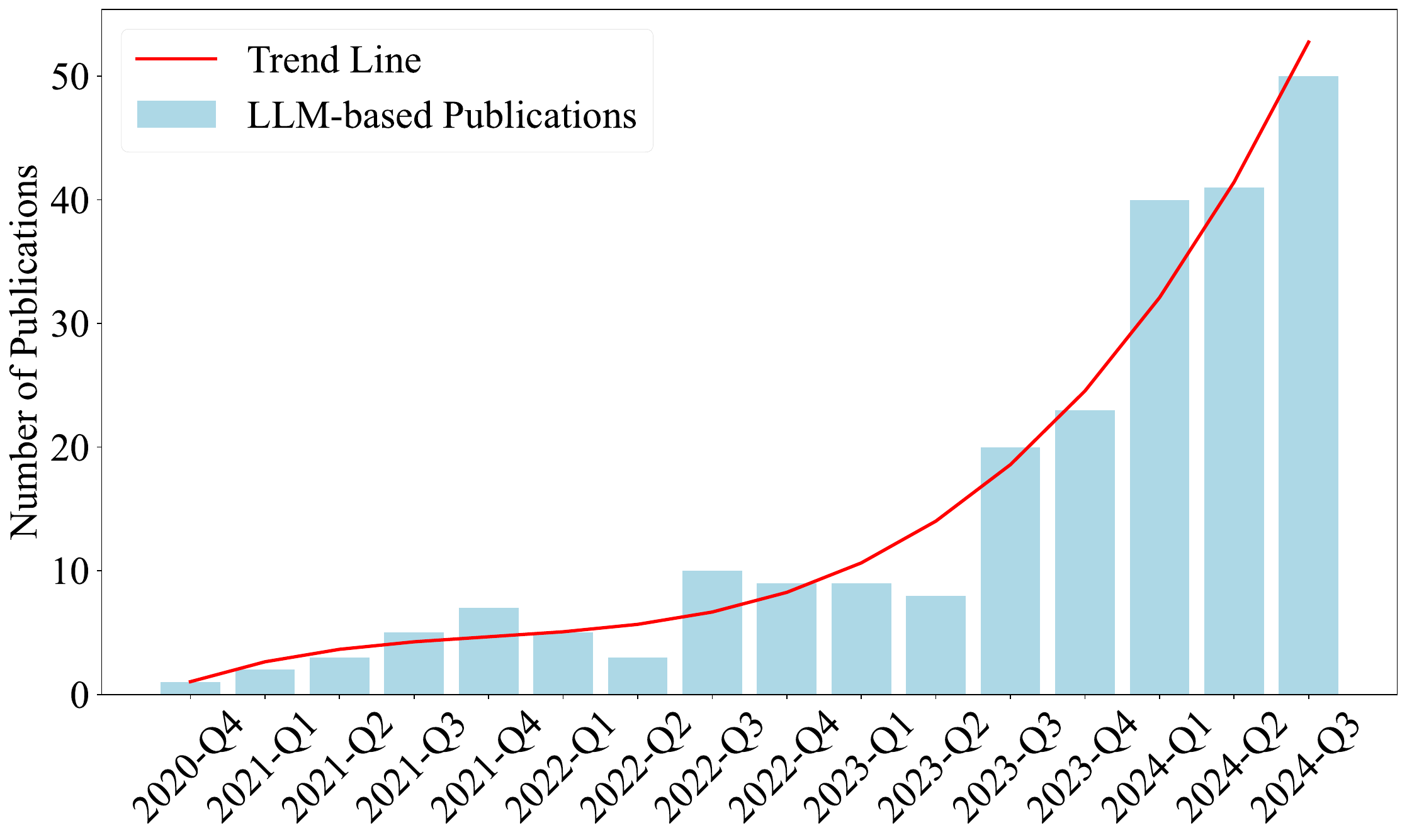}
			\label{fig: llm-ratio}
		\end{minipage}
	}
	\hspace{0.01\linewidth}
	\subfigure[Number of publications in the field of LLM-based AIOps and LLMs (the data for "\# Publications in LLMs" is extended based on~\cite{zhao2023survey})]{
		\begin{minipage}{0.46\linewidth}
			\centering
			\includegraphics[width=\textwidth]{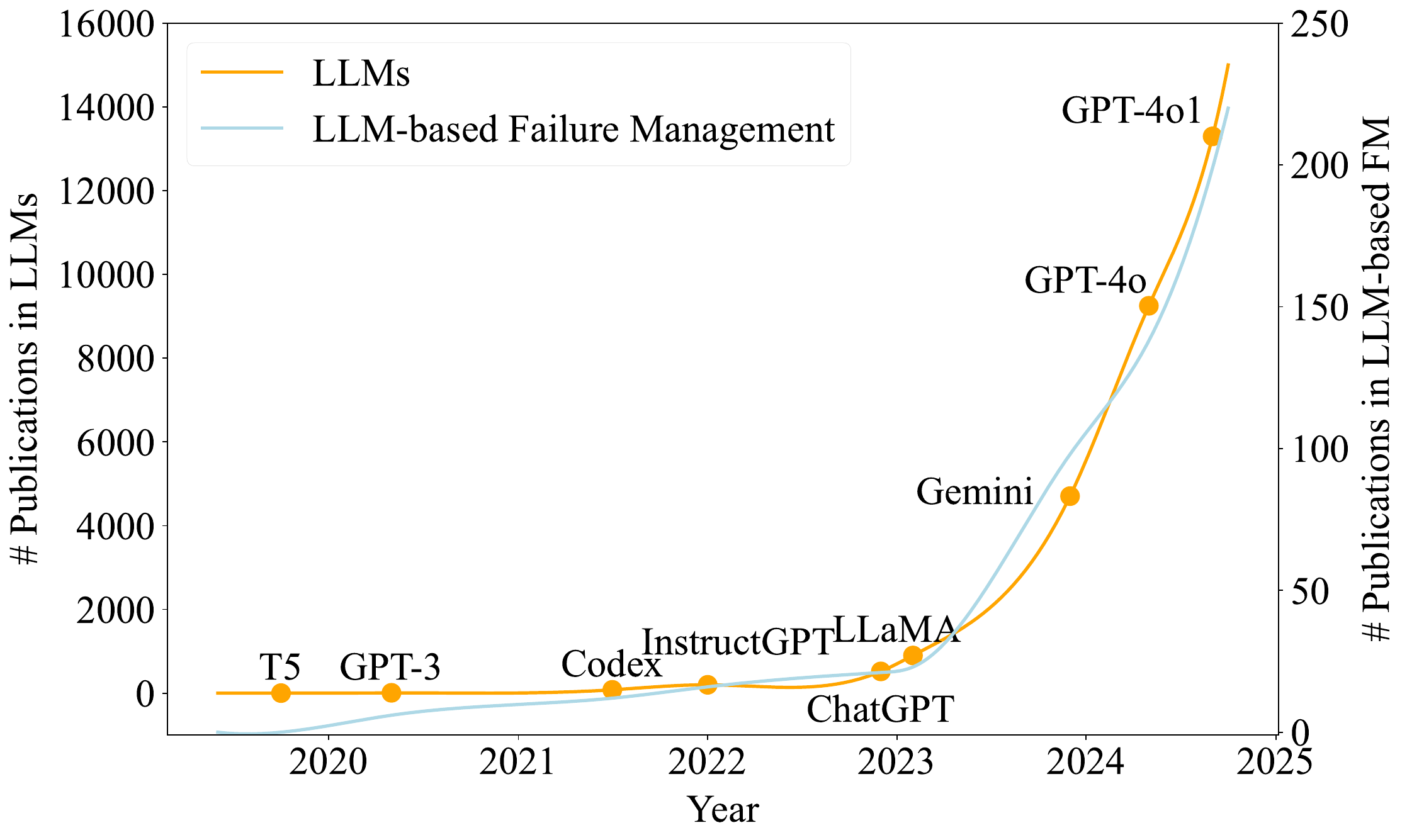}
			\label{fig: paper-num-arxiv}
		\end{minipage}
	}
	\caption{Analysis of Publication Trends in LLM-Based AIOps}
	\label{fig: paper-num}
\end{figure}

Due to the advantages of large language models, an increasing number of efforts in the field of AIOps are now leveraging LLMs, as shown in Figure~\ref{fig: llm-ratio}. Furthermore, the rapid development of LLMs has spurred a thriving research landscape in LLM-based AIOps, as illustrated in Figure~\ref{fig: paper-num-arxiv}. Notably, with the advent of ChatGPT, research on LLM-based AIOps has experienced a significant surge in interest. In summary, as time progresses, the number of works utilizing LLMs continues to grow, and the pace of growth is expected to accelerate in the near future. Therefore, a comprehensive survey analyzing the evolution of AIOps in the era of large models is needed.

In fact, in recent years, numerous literature reviews have summarized research on AIOps. As shown in Table~\ref{tab: survey-comparison}, these works are either based on traditional machine learning or deep learning algorithms and do not use LLM-based approaches~\cite{notaro2021survey, cheng2023ai, remil2024aiops, wei2024log}, or they do not provide a systematic summary of all tasks involved in the full process of AIOps~\cite{su2024large}.

\begin{table}[htbp]
	\centering
	\caption{Comparison of existing AIOps surveys}
	\label{tab: survey-comparison}
	\begin{tabular}{ccp{7.6cm}<{\raggedright\arraybackslash}c}
		\toprule
		Reference & Year & Scope of AIOps Tasks & LLM-based \\
		\midrule
		Paolop et al.~\cite{notaro2021survey} & 2021 & Failure Perception; Root Cause Analysis; Remediation &  \\
		\midrule
		Qian et al.~\cite{cheng2023ai} & 2023 & Incident Detection; Failure Prediction; Root Cause Analysis; Automated Actions &  \\
		\midrule
		Youcef et al.~\cite{remil2024aiops} & 2024 & Incident Reporting; Incident Triage; Incident Diagnosis; Incident Mitigation &  \\
		\midrule
		Wei et al.~\cite{wei2024log} & 2024 & Anomaly Detection &  \\
		\midrule
		Jing et al.~\cite{su2024large} & 2024 & Time-series Forecasting; Anomaly Detection & $\checkmark$ \\
		\midrule
		\textbf{Our work} & - & Data Preprocessing; Failure Perception; Root Cause Analysis; Auto Remediation & $\checkmark$ \\
		\bottomrule
	\end{tabular}
\end{table}

In summary, comprehensive studies exploring AIOps in the era of LLMs are lacking. In this study, we present the first comprehensive survey that covers the entire process of AIOps in the context of large language models. This survey aims to provide researchers with an in-depth understanding of LLM-based approaches for AIOps, facilitating comparisons and contrasts among different methods. It also guides users interested in applying LLM-based AIOps methods by helping them choose suitable algorithms for different application scenarios.

\subsection{Research questions}

With the rapid emergence of numerous LLM-based AIOps solutions—and the strong momentum indicating even greater growth—a comprehensive survey of AIOps in the era of LLMs has become both timely and necessary. The main goal of this survey is to systematically analyze the emerging trends and remaining challenges in LLM-driven AIOps research. By doing so, we aim to offer practical guidance for industrial applications and provide valuable insights to inform future academic research.

\begin{figure}[h]
	\centering
	\includegraphics[width=\textwidth]{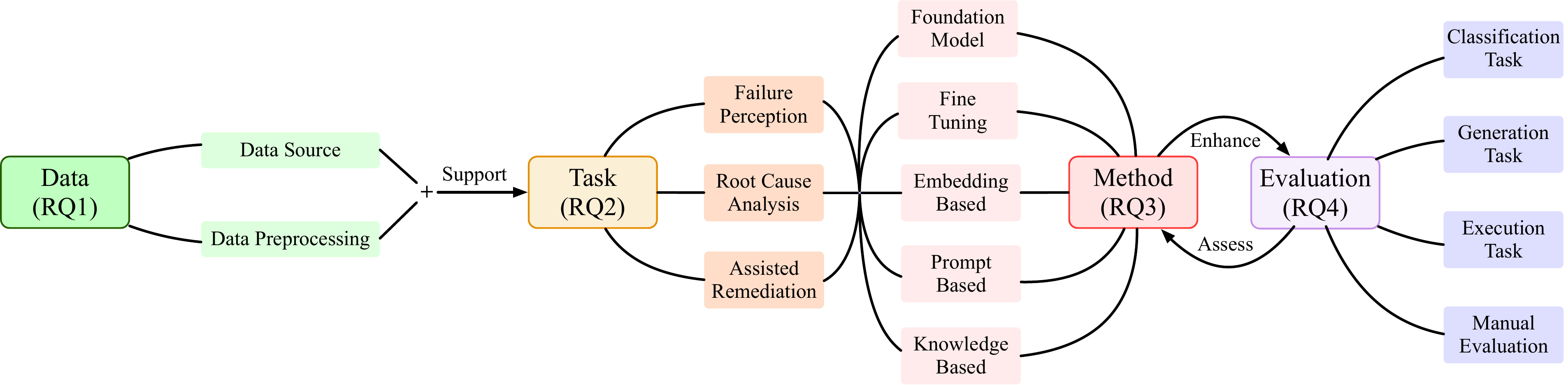}
	\caption{Taxonomy of Research Questions}
	\label{fig: 4rqs}
\end{figure}

To achieve this, we organize our survey around the following research questions, each targeting a critical aspect of how large language models are reshaping AIOps:
\begin{itemize}
	\item \textbf{RQ1:} How has the advent of LLMs transformed the sources and preprocessing methods of data in AIOps?
	\item \textbf{RQ2:} How have the tasks of AIOps evolved with the advent of LLMs?
	\item \textbf{RQ3:} What methods leveraging LLMs are employed in AIOps?
	\item \textbf{RQ4:} How have evaluation methodologies in AIOps adapted to the integration of LLMs?
\end{itemize}
The taxonomy of these research questions is illustrated in Figure~\ref{fig: 4rqs}. Specifically, RQ1 focuses on data sources and preprocessing techniques, which serve as the foundation for the various AIOps tasks examined in RQ2. These tasks are supported by the LLM-based methods reviewed in RQ3. Finally, RQ4 discusses how these methods are assessed through different evaluation strategies. Notably, LLM-based methods may also drive improvements in evaluation itself, forming a feedback loop that continually enhances AIOps performance.

\subsection{Structure of the paper}

This reminder of this survey is structured as follows: In Section 2, we outline the methods used in the systematic review and explain the search strategy. Sections 3 through 6 present the findings relevant to each of the four research questions. Section 7 discusses ongoing challenges and potential future research directions in LLM-based AIOps. Finally, the survey concludes in Section 8.

\section{Systematic Review Process}

To conduct a systematic review, we first establish a research protocol that includes the following steps: defining the search strategy and scope, setting inclusion and exclusion criteria, and presenting an overview of the selected publications~\cite{khan2003five, kitchenham2009systematic}. The purpose of this research protocol is to ensure the study is conducted in a clear, rigorous, and reproducible manner.

\subsection{Search strategy}

To manage the vast body of literature efficiently, we limit our search to five databases\footnote{https://www.scopus.com/, https://www.webofscience.com/, https://ieeexplore.ieee.org/, https://dl.acm.org/, https://arxiv.org/}. Scopus and Web of Science are included due to their comprehensive indexing of peer-reviewed journals and their widespread use in systematic reviews. Additionally, we select IEEE Xplore and the ACM Digital Library, as they are frequently cited in related literature surveys and cover a broad range of applied research fields. Finally, we include arXiv, given its prominence as the leading pre-print platform in the field of computer science.

\begin{figure}[h]
	\centering
	\includegraphics[width=\textwidth]{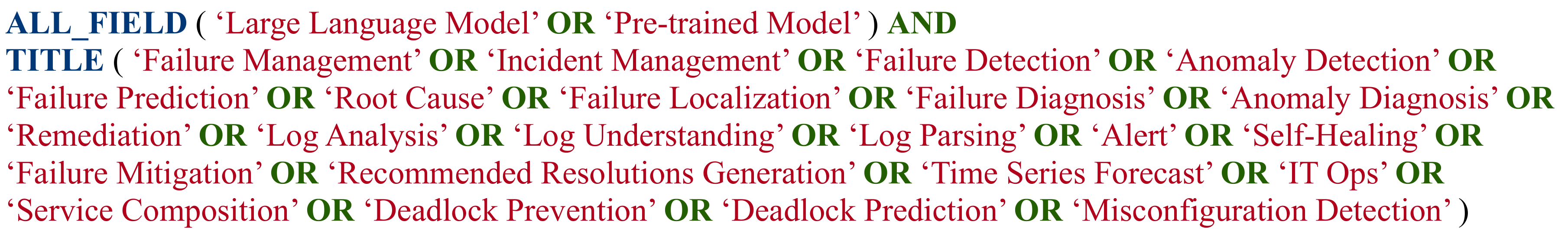}
	\caption{Search Strategy Utilized to Identify Studies on AIOps}
	\label{fig: search}
\end{figure}

Based on our preliminary investigation, we identify relevant keywords for searching literature on LLM-based AIOps and limited our search to papers matching the search string shown in Figure~\ref{fig: search} and focus on works published after 2020. To ensure a comprehensive collection of papers, we include broader keywords during the search, such as \textit{IT Operations} and \textit{Pre-trained Model}. Subsequently, we manually filter out works that fall outside the scope of this survey using exclusion criteria.

\subsection{Selection criteria}

The criteria for selecting papers were as follows. Papers are included if they were deemed relevant to at least one of the proposed research questions. Specifically, a paper is selected if it addresses any of the following: introducing new data sources (RQ1), proposing novel LLM-based methods (RQ2), discussing emerging trends in AIOps tasks (RQ3), or introducing new evaluation metrics or datasets (RQ4). To ensure relevance, papers have to satisfy at least one of the following inclusion criteria:

\begin{itemize} 
	\item \textbf{IC1:} The paper introduces new data sources for AIOps.
	\item \textbf{IC2:} The paper presents novel LLM-based methods for AIOps.
	\item \textbf{IC3:} The paper discusses new changes in AIOps tasks in the LLM era.
	\item \textbf{IC4:} The paper proposes new evaluation metrics or datasets tailored for LLM-based AIOps.
\end{itemize}

Judgments are primarily made based on titles and abstracts. In cases of uncertainty, the full paper is reviewed. To exclude irrelevant works, we apply a series of exclusion criteria to the included papers. Papers are removed if they meet any of the following conditions:

\begin{itemize} 
	\item \textbf{EC1:} The study uses a model with a parameter size smaller than 1 billion.
	\item \textbf{EC2:} The study is unrelated to software systems.
	\item \textbf{EC3:} The study is outside the scope of AIOps.
	\item \textbf{EC4:} The study has been retracted or the full text is unavailable.
	\item \textbf{EC5:} The study only presents conceptual ideas without experimental results.
\end{itemize}

\begin{figure}[h]
	\centering
	\includegraphics[width=\textwidth]{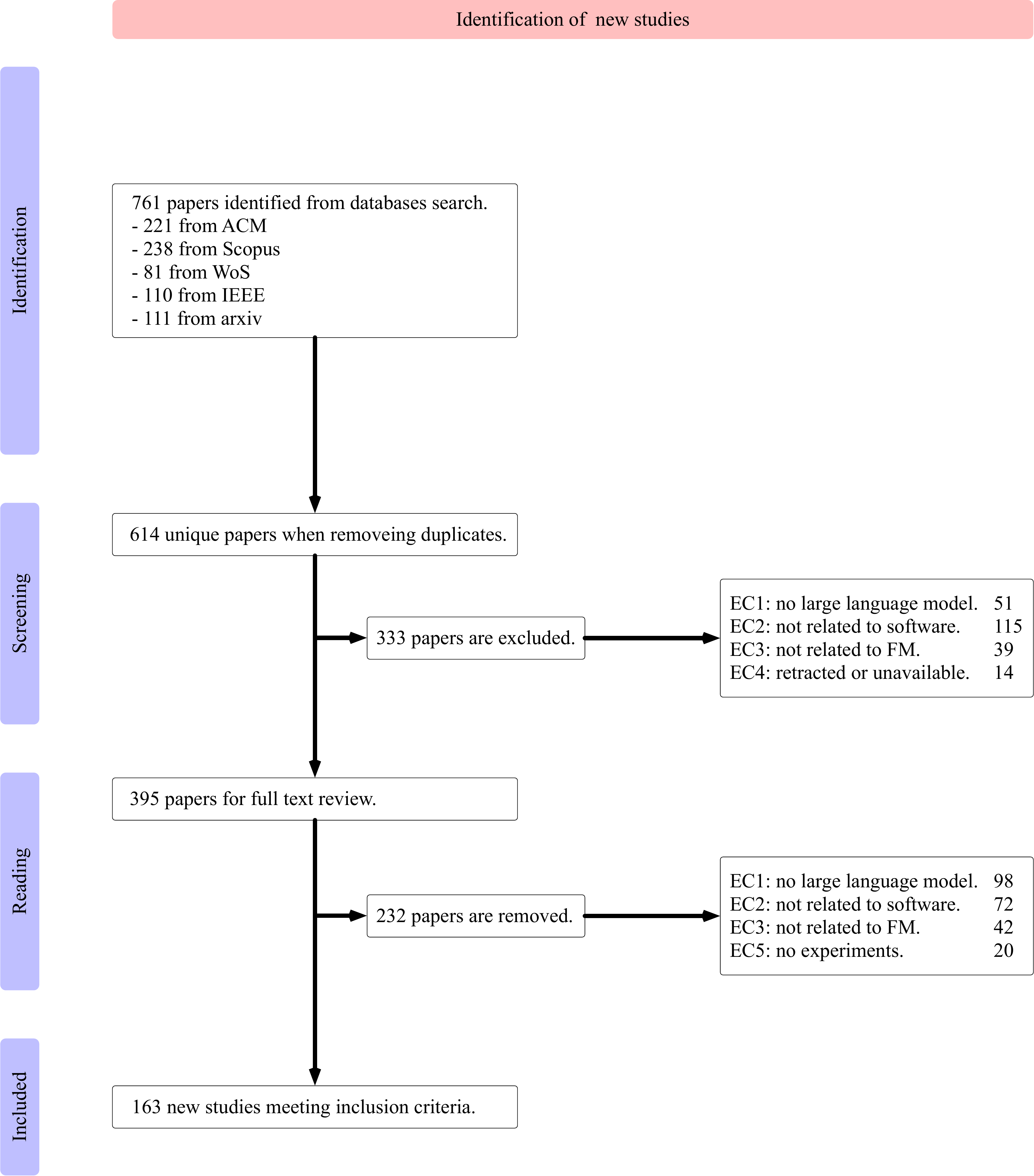}
	\caption{Overview of Paper Selection Procedure}
	\label{fig: selected-publications}
\end{figure}

Each record is reviewed by a primary reviewer, with non-trivial cases discussed and resolved with at least one secondary reviewer. The final set of selected papers is validated by all co-authors. The exclusion criteria are established as follows: Papers have to demonstrate the utilization of large language models. Studies employing smaller models, particularly BERT-based ones, are excluded, as BERT is not widely considered an LLM. Additionally, papers are required to focus on software systems, resulting in the exclusion of studies from other domains, particularly those related to robotics or mechanical devices. Given the broad scope of our search strategy, which included keywords such as IT Operations, we manually filter out papers outside the domain of AIOps. Finally, EC4 and EC5 are used to eliminate low-quality studies, such as retracted papers or those lacking experimental results.

\subsection{Overview of selected publications}

We then present an overview of the workflow for selecting the papers and summarize the corresponding results. As shown in Figure~\ref{fig: selected-publications}, we initially retrieve 761 papers from five databases. After removing duplicates, 614 papers remain. During the screening phase, we exclude 333 papers based on their titles and abstracts using criteria EC1-EC4. This step primarily filter out papers unrelated to AIOps. The remaining 395 papers are downloaded for detailed reading, where we find that 222 papers do not align with the scope of our study. These papers predominantly utilize small-scale models that do not meet our requirements. Ultimately, 163 new studies satisfy our predefined inclusion criteria and are selected for further analysis in this systematic review.

\section{RQ1: Transformations in Data with LLM Integration}
\label{sec:rq1}

In this section, we examine the data sources employed in various studies to explore how data has transformed in the era of LLMs. First, we analyze the new techniques applied to process traditional data sources. Then, we discuss the novel data sources introduced for AIOps in the LLM era.

\subsection{Advancing Preprocessing Techniques for Traditional Data Sources}

To begin, we review the data sources traditionally employed in ML-based and DL-based AIOps~\cite{li2021opengauss, zhou2021dbmind, zhou2018fault, sui2023logkg, yuan2019approach, lin2016log, zhang2022deeptralog, zeng2023traceark, zhang2022putracead, zhang2024reducing, zhang2024towards}. These studies primarily rely on runtime data generated by the system, which encapsulates rich intrinsic and extrinsic information about the operation of software systems. This data can be categorized into three main types: \textbf{metrics}, \textbf{logs}, and \textbf{traces}.

\begin{itemize}
	\item Metrics are quantitative measurements collected from various components of the IT infrastructure, such as CPU usage, memory utilization, disk I/O, network latency, and throughput. These metrics provide real-time performance insights and are critical for monitoring the system's health and performance.
	\item Logs are detailed records of events occurring within the system. They encompass error messages, transaction records, user activities, and system operations. Logs are essential for diagnosing issues, analyzing system behavior, and tracking historical changes over time.
	\item Traces capture the sequence of operations or transactions that a request undergoes in a distributed system. They offer a high-level perspective on the interactions between services, aiding in the identification of performance bottlenecks, service dependencies, and root causes of issues within microservices architectures.
\end{itemize}

In metrics data, a significant challenge stems from the high dimensionality and granularity of data—potentially thousands of indicators generating data points every second. During transmission or storage, data loss is common, and such missing values do not necessarily indicate anomalies in the system. Therefore, imputing missing data is critical, as it can significantly improve the performance of downstream tasks. While many LLM-based metrics methods address imputation as part of broader objectives such as anomaly detection, several studies specifically target the imputation problem. For example, GatGPT~\cite{chen2023gatgpt} employs a graph attention network to pre-train a large language model tailored for metrics imputation. Jacobsen et al.~\cite{jacobsen2024imputation} apply parameter-efficient fine-tuning techniques to adapt LLMs for this task. CRILM~\cite{hayat2025context} takes a different approach by using LLMs to generate contextually informed descriptors for missing values, rather than relying on numerical interpolation.

In trace data, missing or incomplete traces may result from failures in the monitoring infrastructure. To address this, trace generation has emerged as a promising new direction. Although LLM-based research in this area is still limited, Kim et al.~\cite{kim2024large} propose a pioneering approach that fine-tunes a large language model to synthesize realistic workload traces, particularly in the form of microservice call graphs.

Among the various data sources in AIOps, log data has received the most attention in preprocessing research. This is largely due to its unstructured nature, high information density, and critical role in capturing system behaviors. Within this area, log parsing has emerged as the most actively studied task. A significant number of studies leverage LLMs to convert raw log messages into structured representations, which serve as essential inputs for downstream tasks such as failure detection and root cause analysis.

\begin{figure}[h]
	\centering
	\includegraphics[width=\textwidth]{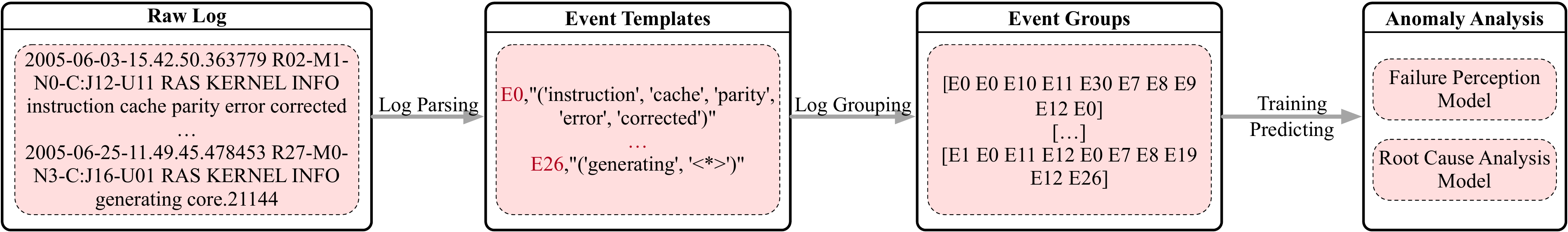}
	\caption{Log-based Failure Perception and Root Cause Analysis: The Common Workflow}
	\label{fig: common-workflow}
\end{figure}

As shown in Figure~\ref{fig: common-workflow}, raw logs consist of semi-structured text encompassing various fields like timestamps and severity levels. For the benefit of downstream tasks, log parsing is employed to transform each log message into a distinct event template, which includes a constant part paired with variable parameters. For example, the log template "E0,('instruction', 'cache', 'parity', 'error', 'corrected')" can be extracted from the log message “2005-06-03-15.42.50.363779 R02-M1-N0-C:J12-U11 RAS KERNEL INFO instruction cache parity error corrected" in Figure~\ref{fig: common-workflow}. After being parsed into event templates, log data can be organized into sequence groups using session, sliding, or fixed windows. Following this, failure perception and root cause analysis are performed on each event group to determine if a failure exists, and if so, to conduct the corresponding root cause analysis.

A common issue in traditional log parsing methods is their lack of generalizability. These methods often rely on manually designed rules or are trained on limited datasets. As a result, their effectiveness significantly decreases when applied to different software systems or when there are changes in log generation rules. The emergence of powerful LLMs, which possess extensive pre-trained knowledge related to code and logging, offers a promising avenue for log parsing. However, the lack of specialized log parsing capabilities currently hinders the accuracy of LLMs in this task. Additionally, the inherent inconsistencies in their responses and the substantial computational overhead prevent the practical adoption of LLM-based log parsing. Consequently, many studies have explored the effectiveness of LLMs for log parsing and have proposed methods to effectively leverage LLMs in this domain.

\textbf{Empirical Study.} Priyanka et al.~\cite{mudgal2023assessment} conducted a study on zero-shot log parsing using ChatGPT. Their findings indicated that the current version of ChatGPT has limited performance in zero-shot log processing, with issues of response inconsistency and scalability. Le et al.~\cite{le2023log} evaluated ChatGPT's ability to perform log parsing through two research questions: the effectiveness of ChatGPT in parsing logs and its performance with different prompting methods. Their results showed that ChatGPT can achieve promising outcomes in log parsing with appropriate prompts, especially with a few-shot approach. These empirical studies demonstrate that while LLMs have the potential for log parsing, they require effective methods to guide them.

\textbf{Prompt-based Approach.} Some studies have adopted prompt-based methods to guide LLMs for effective log parsing. LILAC~\cite{jiang2024lilac} proposes a practical log parsing framework using LLMs with an adaptive parsing cache. LILAC leverages the in-context learning (ICL) capabilities of LLMs by executing a hierarchical candidate sampling algorithm and selecting high-quality demonstrations. It uses an adaptive parsing cache to store and optimize LLM-generated templates, helping mitigate the inefficiency of LLMs by quickly retrieving previously processed log templates. LLMParser~\cite{ma2024llmparser} employs in-context learning and few-shot tuning methods. This approach is tested on four LLMs: Flan-T5-small, Flan-T5-base, LLaMA-7B, and ChatGLM-6B, across 16 open-source systems, and find that smaller LLMs might be more effective for log parsing tasks than more complex ones. Lemur~\cite{guo2024lemur} introduces a novel sampling method inspired by information entropy, effectively clustering typical logs and using the Chain of Thoughts (CoT) method with LLMs to distinguish parameters from invariant tokens. DivLog~\cite{xu2024divlog} combines Retrieval-Augmented Generation (RAG) and ICL methods for log parsing, sampling a diverse set of offline logs as candidate logs and selecting five suitable template candidates for each target log during the parsing process. Sun et al.~\cite{sun2023design} develop a cloud-native log management platform providing log collection, transmission, storage, and system management features, utilizing GPT-3.5 for few-shot log parsing.

\textbf{Fine-tuning Approach.} Other studies have adopted fine-tuning approaches to train pre-trained models specifically for log parsing. OWL~\cite{guo2023owl} employs supervised fine-tuning and mixture adapter tuning techniques to develop a set of large language models for knowledge querying and log parsing, leveraging the LLaMA model and their OWL-Instruct dataset. LogLM~\cite{liu2024loglm} utilizes instruction-based fine-tuning on LLaMA2-7B with seven Loghub datasets~\cite{he2023loghub}, enabling the base model to acquire log parsing capabilities. Mehrabi et al.~\cite{mehrabi2024the} fine-tune the Mistral-7B-Instruct LLM and demonstrate that a fine-tuned, compact LLM can achieve comparable or even superior results to those of large-scale LLMs in log parsing.

\subsection{Emerging Data Sources in AIOps}

The aforementioned traditional data sources are all auto-generated by system, so in this survey we categorize them as system-generated data. However, with the emergence of LLMs, many approaches have started to incorporate information created by humans to provide auxiliary knowledge for managing software system failures~\cite{bhavya2023exploring, zhang2024automated, goel2024x, cao4741492managing, zhang2024lm, xue2023db, guo2023owl, roy2024exploring, shi2023shellgpt, ahmed2023recommending, jiang2024xpert, hamadanian2023holistic, jin2023assess}. These human-generated data types include \textbf{Software Information}, \textbf{Question \& Answer (QA)}, and \textbf{Incident Reports}.

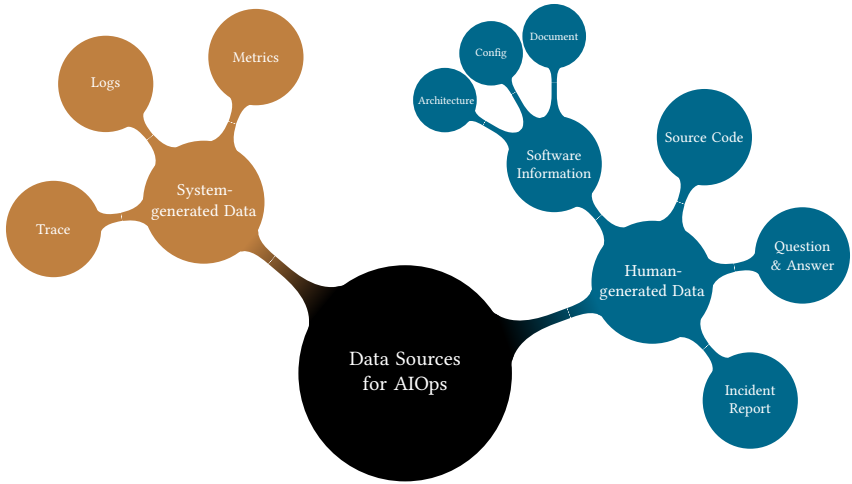
\begin{figure}[htbp]
	\centering
	\begin{tikzpicture}[scale=0.7, transform shape]
		\path[mindmap,concept color=black,text=white]
		node[concept] {Data Sources for AIOps} [clockwise from=120]
		child[concept color=DeepSkyBlue4,grow=20]{
			node[concept] {Human-generated Data} [clockwise from=130]
			child { node[concept] {Software Information} [clockwise from=150]
				child{node[concept] {Architecture}}
				child{node[concept] {Config}}
				child{node[concept] {Document}}
			}
			child { node[concept] {Source Code} }
			child { node[concept] {Question \& Answer}} 
			child { node[concept] {Incident Report} }
		}
		child[concept color=brown,grow=140]{ 
			node[concept] {System-generated Data} [counterclockwise from=70]
			child {node[concept] {Metrics}}
			child {node[concept] {Logs}}
			child {node[concept] {Trace}}
		};
	\end{tikzpicture}
	\caption{Data Source for AIOps in the Era of LLM}
	\label{fig: data-source}
\end{figure}

\textbf{Software Information.} Software information is generated during the software development process and includes details such as \textit{software architecture}, \textit{configurations} and \textit{documentation}. This type of information provides valuable knowledge about the development process, helping AIOps approaches to perform better. By leveraging software information, failure mangement methods can gain insights into the system's design and functionality~\cite{wang2023rcagent, goel2024x, xue2023db, zhou2024llm, zhou2023d, malul2024genkubesec, minna2024analyzing, bhavya2023exploring, li2024llm, shan2024face, shi2023shellgpt, las2024llexus, xie2024cloud, wen2024llm}, which can be crucial for diagnosing issues and implementing effective solutions. For example, Goel et al.~\cite{goel2024x} experimentally demonstrate that incorporating dependent-service descriptions significantly improves the performance of root cause analysis. Moreover, providing service architecture and functionality information helps LLMs more effectively identify the SLO (Service Level Objective) categories of monitors.

\textbf{Source Code.} Source code contains the foundational instructions and logic that define the behavior of software systems. It serves as a critical resource for AIOps by providing deep insights into the system's structure, functionality, and potential areas of vulnerability. In the era of LLMs, source code analysis has been significantly enhanced through automated code comprehension, bug detection, and the generation of potential fixes. For instance, LLMs can analyze source code to identify patterns indicative of bugs or inefficiencies, enabling proactive issue identification. Furthermore, the ability of LLMs to generate code snippets or modifications directly supports tasks such as automated remediation and optimization, bridging the gap between fault diagnosis and corrective action. By leveraging source code, AIOps methods gain a robust foundation for understanding and addressing complex software failures~\cite{wang2023rcagent, khlaisamniang2023generative, othman2023fostering, diaz2024towards, namrud2024kubeplaybook}. For example, RCAgent~\cite{wang2023rcagent} enhances root cause diagnosis for Flink runtime anomalies by jointly leveraging logs and source code. It first uses logs to locate suspicious components and then retrieves the corresponding code snippets. These snippets are subsequently analyzed by an LLM, enabling a deeper semantic understanding of the logs and improving the accuracy of root cause identification.

\textbf{Question \& Answer (QA).} QA data consists of question-and-answer pairs related to operational or developmental knowledge. In the context of LLMs, a rich repository of QA data can serve as a knowledge base, functioning similarly to a search engine, to provide valuable assistance to operations personnel. Additionally, like software information, QA data can offer auxiliary knowledge that enhances AIOps strategies, facilitating quicker and more accurate issue resolution~\cite{liu2023opseval, guo2023owl, zhang2024rag4itops, bhavya2023exploring}. For example, OWL~\cite{guo2023owl} fine-tunes LLaMA2-13B specifically for IT operations-related question answering tasks, and achieves superior performance compared to existing state-of-the-art models.

\textbf{Incident Reports.} Incident reports are often written by software users. When an incident is created, the author specifies a title for the incident and describes relevant details such as error messages, anomalous behavior, and other information that could help with resolution. Before the emergence of LLMs, these incident reports were submitted to on-call engineers (OCEs) for diagnosis. However, due to the powerful natural language processing capabilities of LLMs, many approaches now automatically analyze these incident reports, diagnose faults, and even suggest mitigation steps~\cite{zhang2024automated, goel2024x, zhang2024lm, khlaisamniang2023generative, ahmed2023recommending, jiang2024xpert, kuang2024knowledge, hamadanian2023holistic, jin2023assess, roy2024exploring}. This automation enhances the efficiency of AIOps by quickly identifying and addressing issues based on the detailed information provided in the reports.

In summary, the advent of large language models has significantly expanded the data sources used for AIOps. In addition to traditional system-generated data, a substantial amount of human-generated data with semantic richness but less structured form has been leveraged. These human-generated data types provide valuable context and knowledge that were previously underutilized. By incorporating this new category of data, LLMs enable more comprehensive AIOps approaches, improving the system's ability to detect, diagnose, and remediate failures more effectively.

\section{RQ2: Evolving Tasks in AIOps with LLMs}
\label{sec:rq2}

\begin{figure}[htbp]
	\centering
	\tikzset{
		my node/.style={
			draw,
			align=center,
			thin,
			text width=1.2cm, 
			rounded corners=3,
		},
		my leaf/.style={
			draw,
			align=center,
			thin,
			text width=8.5cm, 
			rounded corners=3,
		}
	}
	\forestset{
		every leaf node/.style={
			if n children=0{#1}{}
		},
		every tree node/.style={
			if n children=0{minimum width=1em}{#1}
		},
	}
	\begin{forest}
		nonleaf/.style={font=\bfseries\scriptsize},
		for tree={%
			every leaf node={my leaf, font=\scriptsize},
			every tree node={my node, font=\scriptsize, l sep-=4.5pt, l-=1.pt},
			anchor=west,
			inner sep=2pt,
			l sep=10pt, 
			s sep=3pt, 
			fit=tight,
			grow'=east,
			edge={ultra thin},
			parent anchor=east,
			child anchor=west,
			if n children=0{}{nonleaf}, 
			edge path={
				\noexpand\path [draw, \forestoption{edge}] (!u.parent anchor) -- +(5pt,0) |- (.child anchor)\forestoption{edge label};
			},
			if={isodd(n_children())}{
				for children={
					if={equal(n,(n_children("!u")+1)/2)}{calign with current}{}
				}
			}{}
		}
		[LLM-based Tasks \\for AIOps, draw=gray, fill=gray!15, text width=2.8cm, text=black
		[Failure Perception, color=lightgreen, fill=lightgreen!15, text width=3cm, text=black
		[Failure Prevention, color=lightgreen, fill=lightgreen!15, text width=5cm, text=black]
		[Failure Prediction, color=lightgreen, fill=lightgreen!15, text width=5cm, text=black]
		[Anomaly Detection, color=lightgreen, fill=lightgreen!15, text width=5cm, text=black]
		]
		[Root Cause Analysis, color=harvestgold, fill=harvestgold!15, text width=3cm, text=black
		[Failure Localization, color=harvestgold, fill=harvestgold!15, text width=5cm, text=black]
		[Failure Category Classification, color=harvestgold, fill=harvestgold!15, text width=5cm, text=black],
		[Root Cause Report Generation *, color=harvestgold, fill=harvestgold!15, text width=5cm, text=black]
		]
		[Assisted Remediation, color=carminepink, fill=carminepink!15, text width=3cm, text=black
		[Assisted Questioning *, color=carminepink, fill=carminepink!15, text width=5cm, text=black]
		[Mitigation Solution Generation, color=carminepink, fill=carminepink!15, text width=5cm, text=black],
		[Command Recommendation *, color=carminepink, fill=carminepink!15, text width=5cm, text=black]
		[Script Generation *, color=carminepink, fill=carminepink!15, text width=5cm, text=black]
		[Automatic Execution *, color=carminepink, fill=carminepink!15, text width=5cm, text=black]
		]
		]
	\end{forest}
	\caption{AIOps Tasks in the Era of LLM (Tasks marked with * are new tasks that have emerged in the era of LLMs)}
	\label{fig: tasks}
\end{figure}
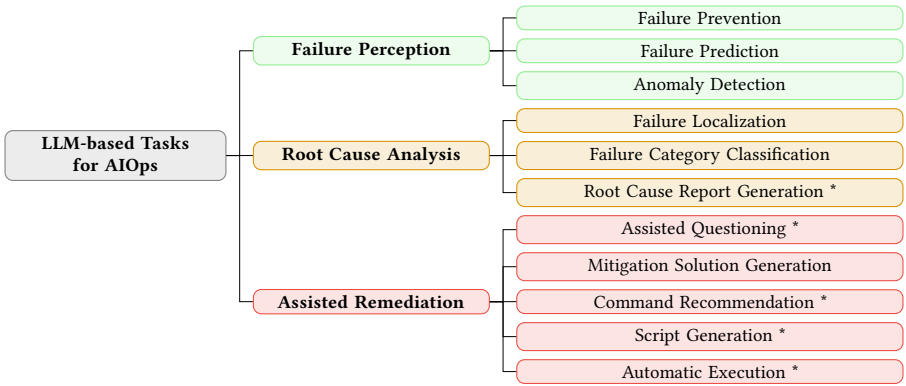

In this section, we present a comprehensive overview of how AIOps tasks have evolved in the era of large language models. We examine the full spectrum of tasks, encompassing both traditional and newly emerging ones, to highlight the transformative role LLMs play in this domain.

AIOps is a multifaceted process comprising several interconnected subtasks. As depicted in Figure~\ref{fig: tasks}, the workflow is divided into three primary stages: failure perception, root cause analysis, and assisted remediation. These stages are sequentially linked as follows:

\textbf{Failure Perception:} This stage focuses on detecting whether anomalies have occurred in the system.

\textbf{Root Cause Analysis:} Once an anomaly is detected, this stage identifies the location and nature of the issue through automated analysis.

\textbf{Assisted Remediation:} Building on the findings of root cause analysis, this stage employs appropriate methods to assist human operators in mitigating the problem effectively.

With the integration of large language models, numerous new subtasks have emerged within these tasks, while many traditional subtasks have undergone significant transformations.

\subsection{Failure Perception}

Failure perception is the foundational step in AIOps, playing a crucial role in the early detection of potential issues to enable proactive measures that prevent failures~\cite{pang2021deep}. This process encompasses several subtasks, including failure prevention, failure prediction, and anomaly detection—all of which existed prior to the advent of large language models. These tasks primarily rely on logs and metrics as data sources, though a few recent studies have incorporated software-related information, particularly configuration data~\cite{malul2024genkubesec, wen2024llm, minna2024analyzing}.

\textbf{Failure Prevention.} This task involves techniques such as software defect prediction and fault injection, which aim to assist in building robust systems and processes to minimize the likelihood of failures~\cite{notaro2021survey}. By identifying potential weaknesses and simulating failure scenarios, these methods help developers design systems with enhanced reliability and fault tolerance. However, in the era of large language models, there has been little work directly addressing failure prevention. The only related study, FAIL~\cite{anandayuvaraj2024fail}, uses LLMs to analyze news articles to preemptively address dependency issues, thereby contributing to failure prevention. This approach differs from traditional failure prevention tasks in its methodology and focus.

\textbf{Failure Prediction.} This task focuses on forecasting potential system failures before they occur by analyzing historical data and identifying patterns that typically precede such failures. This proactive approach alerts maintenance teams to emerging issues, enabling them to implement early remediation measures. However, LLM-based approaches in this domain remain limited. This scarcity arises from the inherent challenge that many failures occur without clear precursors, resulting in methodologies that are either restricted to addressing a narrow range of anomalies or exhibit high rates of missed detections (false negatives). Only a few studies have attempted to leverage LLMs to enhance the effectiveness of failure prediction models~\cite{alharthi2022clairvoyant, alharthi2023time, yang2023diffusion, xiong2023can}, but the tasks they address in failure prediction are constrained in scope.

\textbf{Anomaly Detection.} Due to the inherent limitations of failure prediction, anomaly detection has emerged as a primary focus within failure perception. Its objective is to identify abnormal behaviors or patterns that deviate from the system's normal operations, serving as early indicators of potential issues or failures. This subtask has consistently been the most prominent area of research in failure perception, both before and after the advent of LLMs. With the integration of LLMs, recent efforts have shifted toward enhancing model generalization~\cite{garza2023timegpt, rasul2023lag, dasdecoder, ekambaram2024ttms, chang2023llm4ts, khanal2024domain, hadadi2024anomaly, liu2024anomalyllm, liu2024loglm, le2024prelog, ma2024luk, jin2024large, malul2024genkubesec, sun2024art, yu2024pre, yan2024log} (e.g., developing or fine-tuning foundation models for time series and logs), leveraging large models to improve the performance of smaller ones~\cite{le2021log, zhang2024multivariate, zhang2022logst, ji2023log, setu2024optimizing, tulczyjew2024llmcap, zhang2024end, fariha2024log} (e.g., using LLMs to generate embeddings for logs), and circumventing model training entirely~\cite{xue2023promptcast, jin2023time, cao2023tempo, liu2024lstprompt, zhang2023logprompt, mudgal2023assessment, pan2023raglog, qi2023loggpt, sun2023test, alnegheimish2024large, hegselmann2023tabllm, liu2024interpretable, zhang2024lograg, cui2024logeval, russell2024aad, sun2024semirald, jin2024large, yu2024monitorassistant, shetty2024building, zhong2024dbprompt, hrusto2024autonomous, chatzigeorgakidis2024multicast, zhang2025xraglog, llmelog, eagerlog} (e.g., directly using prompts to predict the next metric or log entry).

\subsection{Root Cause Analysis}

Once anomalies are detected in a software system, it becomes crucial to identify the affected component and the specific nature of the anomaly~\cite{soldani2022anomaly}. Accurate root cause analysis is vital in assisting operations personnel during the repair process. Based on the specific task at hand, these approaches can be categorized into failure localization, failure category classification, and root cause report generation. While failure localization and failure category classification are traditional tasks, root cause report generation is a new task that has emerged with the integration of LLMs. In the LLM era, this task is often addressed using multimodal data within a single workflow—primarily traces, metrics, and logs. A growing body of work, particularly from Microsoft, focuses on analyzing incident reports in this context~\cite{zhang2024automated, goel2024x, zhang2024lm, khlaisamniang2023generative, ahmed2023recommending, jiang2024xpert, kuang2024knowledge, hamadanian2023holistic, jin2023assess, roy2024exploring}. Additionally, software information and source code are frequently used to complement and enhance the analysis of these primary data sources.

\begin{figure}[h]
	\centering
	\includegraphics[width=\textwidth]{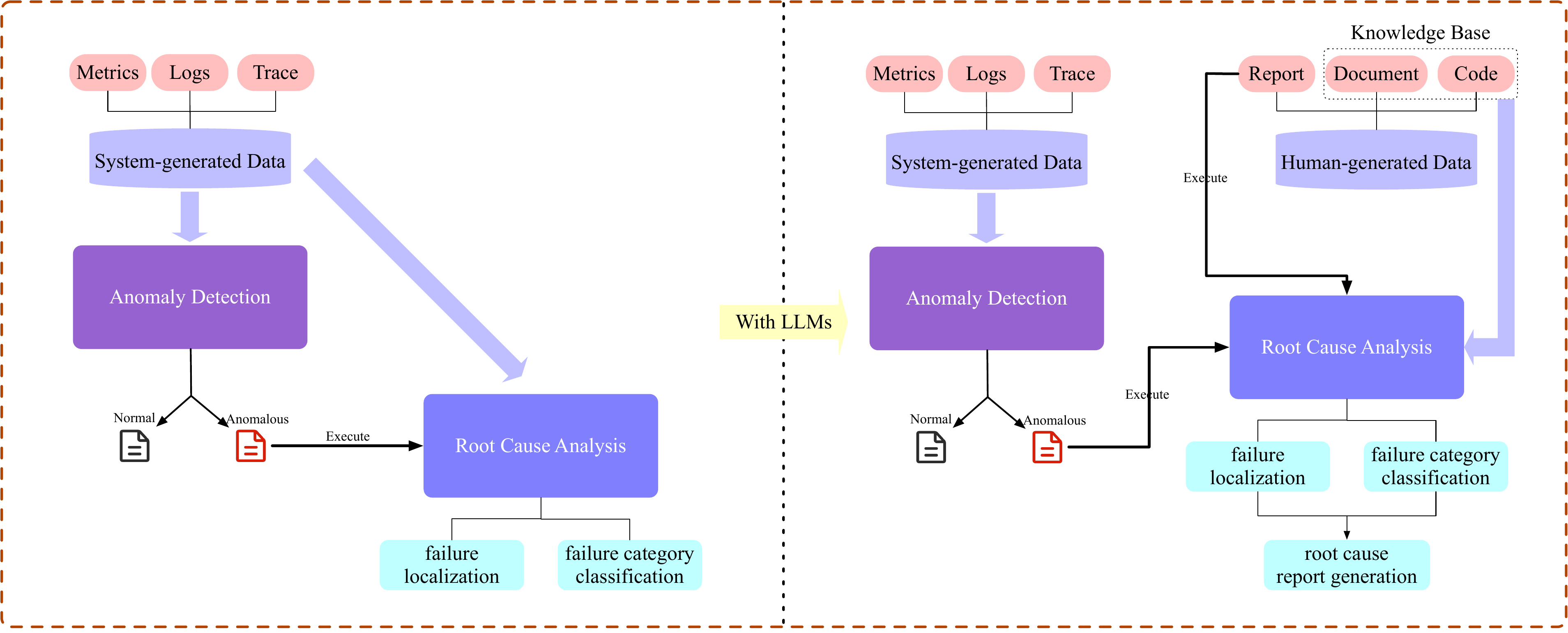}
	\caption{Evolution of Root Cause Analysis with the Rise of LLMs}
	\label{fig: root-cause-analysis}
\end{figure}

As illustrated in Figure~\ref{fig: root-cause-analysis}, before the rise of large language models, these tasks typically relied on system-generated data and utilized automated failure perception methods to identify anomalies, which was then used for failure localization and failure category classification. However, with the advent of large language models, the starting point for root cause analysis has shifted from automated failure perception to user-generated data, particularly incident reports which has introduced the ability to analyze natural language. Additionally, root cause analysis now incorporates other human-generated data, such as documentation and code, as supplementary knowledge sources to enhance the analysis. Furthermore, based on the natural language understanding and generation capabilities of LLMs, it is now possible to bypass the processes of failure localization and failure category classification, and directly generate root cause reports.

\textbf{Failure Localization.} This task aims to identify the specific component or machine where an anomaly occurred, often using methods such as causal discovery. In microservices environments, it helps pinpoint the exact service or machine experiencing the issue. Additionally, it may involve identifying the specific log or metric entry that signals the onset of the anomaly. In the era of LLMs, many studies have focused on leveraging these models to locate erroneous configurations~\cite{shan2024face, malul2024genkubesec, wen2024llm} and identify failure nodes in microservices systems~\cite{sarda2024leveraging, sarda2023adarma, zhang2024mabc, roy2024exploring, shi2024enhancing, xie2024cloud, han2024potential, zhang2025thinkfl}. Some studies have also explored the relationships within various input data sources~\cite{li2024llm, ma2024knowlog, hrusto2024autonomous}.

\textbf{Failure Category Classification.} This task addresses the challenge of determining the type of anomaly the system is experiencing. This task, often formulated as a multi-class classification problem, categorizes anomalies into predefined types, such as CPU resource shortages, memory constraints, or software configuration errors. Traditional failure category classification typically requires predefined failure categories and trains multi-class models strictly according to these categories. As a result, the effectiveness of these models is limited by the predefined categories. However, the advent of large language models has expanded the scope of this task. Many studies now attempt to directly construct appropriate prompts and incorporate external knowledge, enabling LLMs to autonomously identify failure categories~\cite{zhang2024lm, zhou2024llm, chen2024automatic, zhou2023d, quan2023heterogeneous, zhang2024mabc, shetty2024building, zhang2025scalalog, zhang2025agentfm}. Other works have pre-trained foundation models for time-series or log data, allowing the addition of new categories with minimal fine-tuning on the existing model~\cite{zhou2023one, sun2023test, gupta2023learning, ma2024luk, sun2024art, han2024potential}.

\textbf{Root Cause Report Generation.} This task emerges with the enhanced natural language generation and reasoning capabilities of large language models. This development goes beyond isolated tasks such as failure localization and failure category classification. Instead, there is a growing focus on creating comprehensive and more easily understandable root cause reports. These reports not only include information on failure location and category classification but also provide detailed reasoning about the underlying cause of the failure~\cite{roy2024exploring, zhang2024automated, wang2023rcagent, goel2024x, ahmed2023recommending, roy2024exploring, yu2024monitorassistant, hrusto2024autonomous}. This comprehensive approach significantly aids system maintenance personnel in analyzing and resolving issues more efficiently.

\begin{figure}[h]
	\centering
	\includegraphics[width=\textwidth]{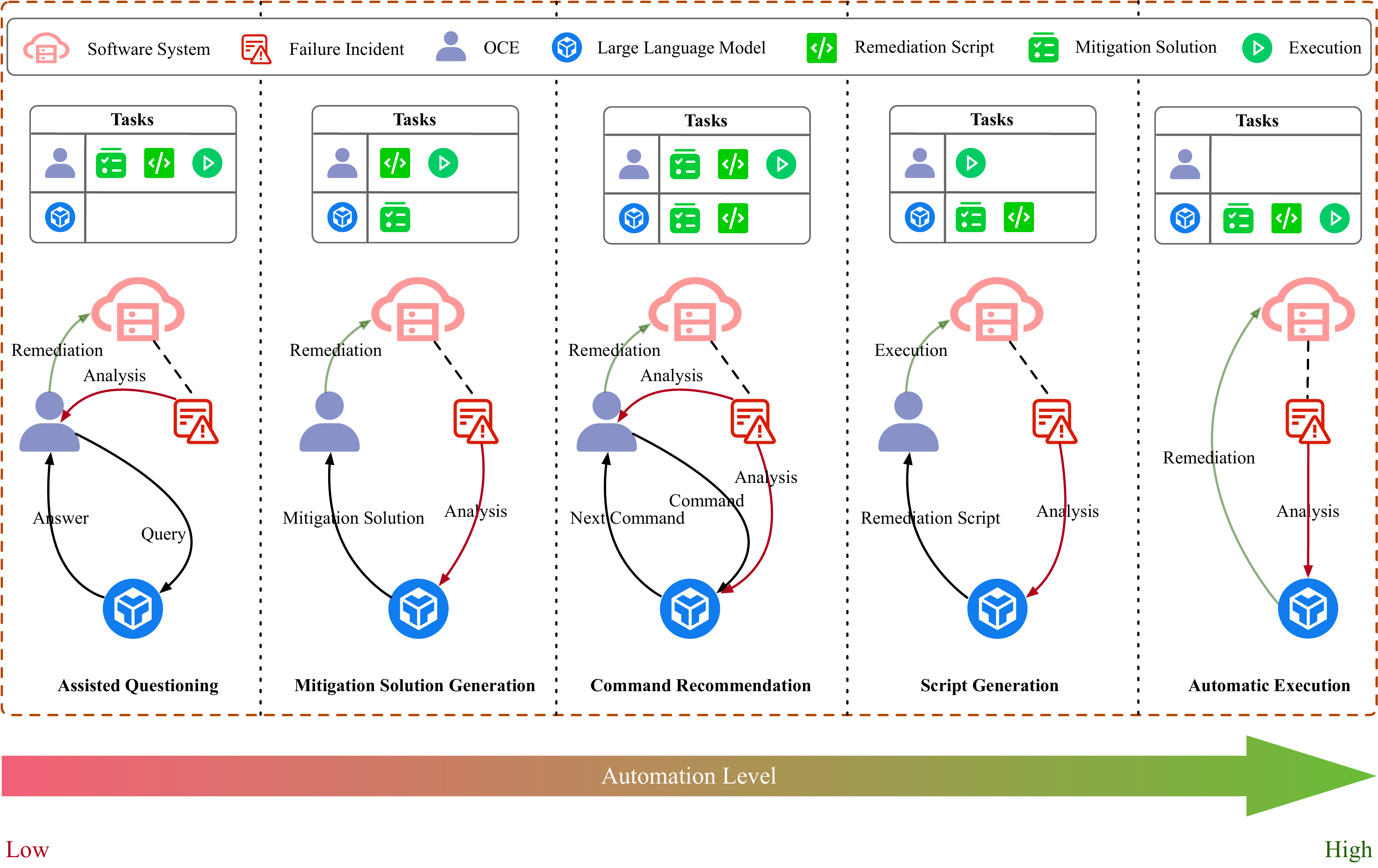}
	\caption{Various Types of Auto Remediation Approaches}
	\label{fig: auto-remediation}
\end{figure}

\subsection{Assisted Remediation}

After identifying the type and location of the software system anomaly, the next step is to automatically mitigate and resolve the issue based on this information~\cite{cheng2023ai}. Before the advent of large language models, the automation level of these remediation tasks was relatively low. However, with the introduction of LLMs, the degree of automation has significantly improved. As illustrated in Figure~\ref{fig: auto-remediation}, this survey classifies assisted remediation approaches by automation level into five categories, arranged by increasing levels of automation: Assisted Questioning, Mitigation Solution Generation, Command Recommendation, Script Generation, and Automatic Execution. Among these, only Mitigation Solution Generation existed in traditional assisted remediation approaches, while the other tasks have emerged as new possibilities enabled by LLMs. These automation levels were determined through a combination of literature review and expert interviews. First, we summarized the entities involved in AIOps based on relevant papers published by the Microsoft research team~\cite{chenaiopslab, shetty2024building, zhang2024enhanced, sun2024art, yu2024monitorassistant, zhang2024automated, zhang2024failure, chen2024automatic, jiang2024xpert}. Using these entities, we identify and categorize the relevant literature for our review. Subsequently, we conduct interviews with OCEs from partner companies to evaluate and refine our proposed assisted remediation framework. Unlike earlier AIOps tasks, this stage does not rely on a single primary data source. Instead, it integrates all data modalities discussed in RQ1. For example, Assisted Questioning primarily relies on historical QA data; Mitigation Solution Generation may leverage incident reports, traces, metrics, and logs—often enriched with software information used as a knowledge base; Command Recommendation and Script Generation, on the other hand, typically depend on source code and configuration data.

\textbf{Assisted Questioning.} This task involves using LLMs to assist operations personnel by answering system-related questions. By allowing operations staff to directly query the LLM software, detailed responses can be obtained quickly, speeding up the resolution of software system failures~\cite{liu2023opseval, park2023formulating, guo2023owl, zhang2024rag4itops, huang2024gloss}. This type of task emerged with the rise of LLMs, particularly GPT-3.5.

\textbf{Mitigation Solution Generation.} In this task, LLMs generate potential mitigation solutions for the detected anomalies. These solutions are often derived from large datasets of historical incident reports and resolutions, providing actionable suggestions to operations personnel for addressing issues~\cite{goel2024x, wang2023network, ahmed2023recommending, hamadanian2023holistic, liu2024loglm, wang2024logexpert, minna2024analyzing, shetty2024building}. Before LLMs, this subtask primarily involved incident triage~\cite{zeng2017knowledge}, but these methods are not intelligent. Mitigation solution generation appears more advanced and tailored.

\textbf{Command Recommendation.} This task leverages LLMs to recommend pre-existing scripts that can be used to remediate the detected issue. The system can suggest the next command to enter when operations staff input commands (e.g., shell commands), thus shortening the repair time~\cite{xue2023db, shi2023shellgpt}. This method gains popularity with the advent of LLMs.

\textbf{Script Generation.} This task skips the command recommendation step, directly using LLMs to generate custom scripts tailored to resolve specific detected anomalies. This involves creating new scripts based on the details of the issue and the context provided by system logs and metrics, allowing for more precise and effective remediation actions~\cite{sarda2024leveraging, jiang2024xpert, wang2023low, pesl2023uncovering, aiello2023service, malul2024genkubesec, namrud2024kubeplaybook}.

\textbf{Automatic Execution.} The highest level of automation involves LLMs not only generating the necessary remediation scripts but also executing them automatically. This end-to-end process ensures that detected anomalies are addressed without manual intervention, significantly speeding up resolution time and reducing the workload on operations personnel. Although this method is highly attractive, related work is limited, and its practical effectiveness remains to be verified~\cite{cao4741492managing, khlaisamniang2023generative, othman2023fostering, sarda2023adarma, las2024llexus}.

\section{RQ3: LLM-based Methods for AIOps}
\label{sec:rq3}

In this section, we introduce various LLM-based methods and explore the specific domains in which they are frequently applied to different AIOps tasks. Additionally, we analyze the respective advantages and limitations of applying these approaches.

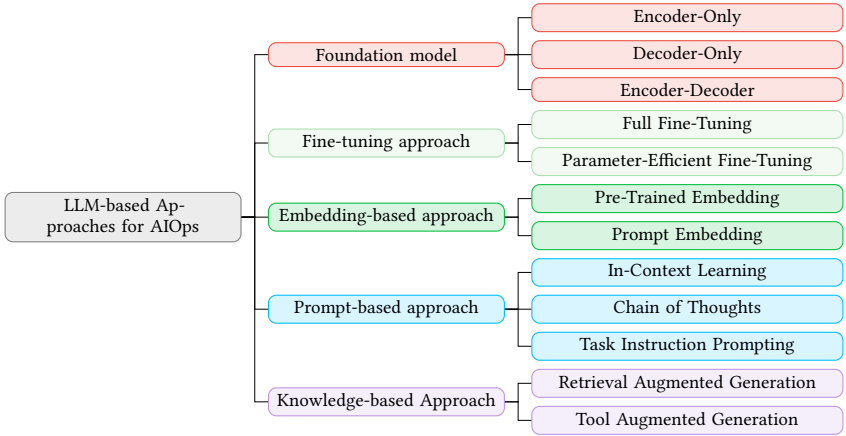
\begin{figure}[h]
	\centering
	\tikzset{
		my node/.style={
			draw,
			align=center,
			thin,
			text width=1.2cm, 
			rounded corners=3,
		},
		my leaf/.style={
			draw,
			align=center,
			thin,
			text width=8.5cm, 
			rounded corners=3,
		}
	}
	\forestset{
		every leaf node/.style={
			if n children=0{#1}{}
		},
		every tree node/.style={
			if n children=0{minimum width=1em}{#1}
		},
	}
	\begin{forest}
		nonleaf/.style={font=\scriptsize},
		for tree={%
			every leaf node={my node, font=\scriptsize, l sep-=4.5pt, l-=1.pt},
			every tree node={my node, font=\scriptsize, l sep-=4.5pt, l-=1.pt},
			anchor=west,
			inner sep=2pt,
			l sep=10pt, 
			s sep=3pt, 
			fit=tight,
			grow'=east,
			edge={ultra thin},
			parent anchor=east,
			child anchor=west,
			if n children=0{}{nonleaf}, 
			edge path={
				\noexpand\path [draw, \forestoption{edge}] (!u.parent anchor) -- +(5pt,0) |- (.child anchor)\forestoption{edge label};
			},
			if={isodd(n_children())}{
				for children={
					if={equal(n,(n_children("!u")+1)/2)}{calign with current}{}
				}
			}{}
		}
		[LLM-based Approaches for AIOps, draw=gray, fill=gray!15, text width=3cm, text=black
		[Foundation model, color=carminepink, fill=carminepink!15, text width=3cm, text=black
		[Encoder-Only, color=carminepink, fill=carminepink!15, text width=4cm, text=black]
		[Decoder-Only, color=carminepink, fill=carminepink!15, text width=4cm, text=black]
		[Encoder-Decoder, color=carminepink, fill=carminepink!15, text width=4cm, text=black]
		]
		[Fine-tuning approach, color=celadon, fill=celadon!15, text width=3cm, text=black
		[Full Fine-Tuning, color=celadon, fill=celadon!15, text width=4cm, text=black]
		[Parameter-Efficient Fine-Tuning , color=celadon, fill=celadon!15, text width=4cm, text=black]
		]
		[Embedding-based approach, color=darkpastelgreen, fill=darkpastelgreen!15, text width=3cm, text=black
		[Pre-Trained Embedding, color=darkpastelgreen, fill=darkpastelgreen!15, text width=4cm, text=black]
		[Prompt Embedding, color=darkpastelgreen, fill=darkpastelgreen!15, text width=4cm, text=black]
		]
		[Prompt-based approach, color=capri, fill=capri!15, text width=3cm, text=black
		[In-Context Learning, color=capri, fill=capri!15, text width=4cm, text=black]
		[Chain of Thoughts, color=capri, fill=capri!15, text width=4cm, text=black],
		[Task Instruction Prompting, color=capri, fill=capri!15, text width=4cm, text=black]
		]
		[Knowledge-based Approach, color=brightlavender, fill=brightlavender!15, text width=3cm, text=black
		[Retrieval Augmented Generation, color=brightlavender, fill=brightlavender!15, text width=4cm, text=black]
		[Tool Augmented Generation, color=brightlavender, fill=brightlavender!15, text width=4cm, text=black]
		]
		]
	\end{forest}
	\caption{LLM-based Approaches for AIOps}
	\label{fig: llm-approach}
\end{figure}

Many surveys have proposed various LLM-based approaches~\cite{li2024pre, chang2024survey, minaee2024large, gu2023systematic, sahoo2024systematic}, many of which have been applied to address AIOps tasks. As shown in Figure~\ref{fig: llm-approach}, in this survey, we categorize these approaches into five groups: foundation model, fine-tuning approach, embedding-based approach, prompt-based approach, and knowledge-based approach. The foundation model refers to pre-trained language models that serve as the base for AIOps tasks without further modifications. The fine-tuning approach involves adapting these pre-trained models to specific tasks through additional training on task-specific data. The embedding-based approach uses the representations generated by pre-trained models to capture semantic information and improve task performance. The prompt-based approach leverages natural language prompts to guide the model's responses, enabling it to perform specific tasks based on the given instructions. The knowledge-based approach integrates external knowledge sources to enhance the model's capabilities by providing relevant context or executing specialized tools for more accurate and informed AIOps task completion.

\subsection{Foundation Model}

As mentioned earlier, in the era of large language models, an increasing number of efforts focus on enhancing model generalization, with one of the most straightforward approaches being the development of foundation models. Currently, almost all large language models are based on the Transformer architecture, with differences arising from the utilization of encoder and decoder blocks~\cite{Ye2024ASO}. Specifically, the encoder maps the input sequence into latent representations, while the decoder processes these representations to generate the output sequence. Consequently, these works can be categorized into three architectural types: \textbf{encoder-only}, \textbf{decoder-only}, and \textbf{encoder-decoder}. In the context of AIOps tasks, such models are predominantly applied in anomaly detection and failure category classification, commonly referred to as metric and log foundation models.

\textbf{Encoder-Only Models.} Encoder-only architectures focus on processing the entire input sequence simultaneously, extracting contextual information for predictions. These models generally have smaller parameter sizes and were more prominent in the earlier stages of large language model development. Despite their reduced popularity in foundation model pretraining, they remain effective for specific applications requiring fine-grained input analysis.

LoFI~\cite{huang2024demystifying} employs an encoder-only design with a prompt-based tuning method to extract detailed insights from log data, particularly for failure diagnosis tasks. Similarly, ART~\cite{sun2024art} uses an encoder-only framework to pretrain a system that integrates anomaly detection, failure triage, and root cause localization, focusing on metrics data. These examples demonstrate the continued relevance of encoder-only models in domains where efficiency and detailed input analysis are prioritized.

\textbf{Decoder-Only Models.} Decoder-only architectures are characterized by autoregressive token generation, where each token depends on previous ones. Popularized by models like GPT-3, these architectures typically involve larger parameter sizes and are well-suited for generative tasks.

Several studies have adapted decoder-only architectures for AIOps. Lag-Llama~\cite{rasul2023lag} pretrains a foundation model for univariate probabilistic time series forecasting using lags as covariates within a decoder-only transformer framework. TimesFM~\cite{dasdecoder} employs a patched-decoder-style attention model to robustly address varying forecasting history lengths, prediction horizons, and temporal granularities. For domain-specific tasks, ShellGPT~\cite{shi2023shellgpt} adapts a GPT architecture to align shell scripting with natural language, embedding domain knowledge for automation tasks. Timer~\cite{liutimer} builds a GPT-style model tailored for long-term sequence modeling (LTSMs), combining forecasting, imputation, and anomaly detection in a unified framework. These works highlight the generative power and flexibility of decoder-only models for complex, sequential tasks.

\textbf{Encoder-Decoder Models.} Encoder-decoder architectures integrate both encoder and decoder blocks, allowing distinct phases for input processing and output generation. This design is particularly well-suited for tasks requiring intricate input-output mappings. Foundational examples such as T5 and BART epitomize this architecture, balancing parameter size for flexibility and efficiency.

In AIOps, only a few studies leverage this architecture. TimeGPT~\cite{liao2024timegpt} employs a deep encoder-decoder structure to pretrain a foundation model for time series data, focusing on extracting complex patterns from over 100 billion data points~\cite{Ye2024ASO}. Similarly, SimMTM~\cite{dong2024simmtm} reconstructs masked time points in metrics data by leveraging the encoder-decoder structure, which aggregates information from neighboring points. This method excels at recovering temporal information outside the manifold. For log analytics, PreLog~\cite{le2024prelog} uses this architecture with entry-level and sequence-level objectives to train a model effective in log parsing and log-based anomaly detection. Moreover, KAD-Disformer~\cite{yu2024pre} adapts an encoder-decoder transformer for disentangling temporal and spatial dynamics in metrics data, enhancing anomaly detection performance. These works highlight the versatility of encoder-decoder models in capturing diverse patterns across time-series and log-based data.

\subsection{Fine-tuning Approach}

Directly applying general foundation models to AIOps tasks often does not yield optimal results. Therefore, many AIOps approaches fine-tune these models using domain-specific datasets. This fine-tuning can be categorized into two types: \textbf{full fine-tuning} and \textbf{parameter-efficient fine-tuning}.Most of these methods primarily rely on metrics and log data. However, a smaller subset of approaches also incorporates QA data and source code during the fine-tuning process.

\textbf{Full Fine-Tuning.} Full fine-tuning updates all parameters of a foundation model with domain-specific data, enabling the model to adapt comprehensively to new tasks. However, due to the computational complexity and large parameter sizes of models like GPT-3.5 and GPT-4, most current efforts focus on relatively smaller models with fewer than 100 billion parameters, such as Codex and LLaMA.

For instance, PromptCast~\cite{xue2023promptcast} fine-tunes text-visible models such as T5 and BART with carefully designed prompts to enable forecasting-based anomaly detection~\cite{Ye2024ASO}. Similarly, UniTime~\cite{liu2024unitime} adopts a traditional fine-tuning approach on GPT-2 to develop a unified model for cross-domain time series forecasting. These works demonstrate the efficacy of full fine-tuning in adapting general-purpose models to time-series-specific tasks.

Other works explore innovative fine-tuning methodologies. AnomalyLLM~\cite{liu2024anomalyllm} leverages LLaMA2-7B and introduces a dynamic-aware contrastive fine-tuning technique to address graph-based anomaly detection. Meanwhile, RAG4ITOps~\cite{zhang2024rag4itops} uses a Qwen-14B-base model and implements Continue Retrieval Augmented Fine-Tuning to build an assisted questioning system tailored to IT operations.

\textbf{Parameter-Efficient Fine-Tuning.} Parameter-efficient fine-tuning offers a more computationally feasible alternative by updating only a subset of model parameters. This approach includes techniques such as layer-freezing, adapter tuning, and task-conditional fine-tuning, which are particularly effective in resource-constrained environments.

Several studies demonstrate the utility of parameter-efficient methods for AIOps. Komal et al.~\cite{sarda2024leveraging} investigate the performance of in-context tuning and few-shot tuning on LLaMA2-70B for auto-remediation tasks, highlighting the adaptability of large language models with minimal updates. Similarly, Toufique et al.~\cite{ahmed2023recommending} conduct a large-scale empirical study on models like GPT-Neo and GPT-2, evaluating their effectiveness in assisting engineers with root cause analysis and production incident mitigation through few-shot tuning.

A common strategy in this domain involves using adapters to improve fine-tuning efficiency. OWL~\cite{guo2023owl} applies a mixture-of-adapters strategy to LLaMA2-13B, achieving significant improvements across tasks like assisted questioning and log-based anomaly detection. Similarly, LLM4TS~\cite{chang2023llm4ts} focuses on freezing the majority of GPT-2 parameters and fine-tuning only a small subset for general forecasting-based anomaly detection.

Instruction tuning has also emerged as a promising parameter-efficient approach. LogLM~\cite{liu2024loglm} fine-tunes LLaMA2-7B using instruction-based tasks, enabling it to handle a wide range of log analysis tasks such as log parsing and anomaly detection. Additionally, models like CrashEventLLM~\cite{mudgal2024crasheventllm}, Fatemeh et al.~\cite{hadadi2024anomaly}, and Hongwei et al.~\cite{jin2024large} explore supervised fine-tuning on models like GPT-3 and LLaMA2-7B to optimize log-based anomaly detection.

Few-shot tuning has been particularly effective for log parsing and analysis. For example, LLMParser~\cite{ma2024llmparser} and Maryam et al.~\cite{mehrabi2024effective} evaluate few-shot tuning on models such as Flan-T5-small, Flan-T5-base, LLaMA-7B, ChatGLM-6B, and Mistral-7B-Instruct, demonstrating the ability of lightweight fine-tuning to adapt general models for specialized tasks.

\subsection{Embedding-based Approach}

Many data sources in AIOps, such as logs and documentation, contain rich semantic information, making embeddings a crucial component in capturing and representing this information in a structured format. Embedding-based approaches for AIOps can be broadly categorized into two types: \textbf{pre-trained embedding} and \textbf{prompt embedding}. Pre-trained embeddings leverage pre-existing large language models (LLMs) to extract semantic representations directly, while prompt embeddings involve designing task-specific prompts to activate the LLMs’ capabilities for specialized data processing such as metric data.

\textbf{Pre-Trained Embedding.} Pre-trained embeddings rely on leveraging representations generated by models like GPT or T5, which have been trained on extensive corpora and can encode rich semantic information. These embeddings are versatile and can be applied to various AIOps tasks without further fine-tuning, making them particularly popular for tasks involving log analysis.

For instance, Harold et al.~\cite{ott2021robust} utilized embeddings from GPT-2, chaining these embeddings through time using a recurrent neural network (BiLSTM) to model and learn normal system behaviors. Similarly, Egil et al.~\cite{karlsen2024large} employed GPT-2-based embeddings within an autoencoding framework, compressing the representations and applying self-supervised learning for log-based anomaly detection. Setu et al.~\cite{setu2024optimizing} focused on GPT-3, utilizing its word embeddings and tokenizer to transform log data into a format suitable for identifying unusual patterns and anomalies effectively.

\textbf{Prompt Embedding.} Prompt embedding works focus on designing specific embedding methods tailored for LLMs to effectively capture semantic information from various data sources in AIOps. By providing natural language prompts or instructions, these methods activate the LLM's processing capability to generate task-specific embeddings. This approach enables flexible and task-specific embeddings suitable for different AIOps applications. Currently, many works based on metrics utilize this approach to transform metrics data into a format more suitable for LLM understanding.

Time-LLM~\cite{jin2023time} proposes Prompt-as-Prefix (PaP), which enriches the input context and directs the transformation of reprogrammed input patches. The transformed time series patches from the LLM are then projected to obtain the forecasts. TEST~\cite{sun2023test} builds an encoder based on metrics data to embed them by instance-wise, feature-wise, and text-prototype-aligned contrast, and then creates prompts to make the LLM more receptive to embeddings, achieving excellent results in failure category classification. Nate et al.~\cite{gruver2024large} used prompt embedding techniques with GPT-3, LLaMA-2, and GPT-4 to accomplish anomaly detection and metrics imputation tasks. TEMPO~\cite{cao2023tempo} decouples complex time series into trend, seasonal, and residual components, mapping them to corresponding latent spaces to create inputs recognizable by GPT. To achieve this, TEMPO constructs a prompt pool and assigns different prompts to different decoupled components, enabling the model to adapt to changes in the time series distribution using historical information.

\subsection{Prompt-based Approach}

In addition to leveraging the embedding capabilities of large language models, many AIOps approaches directly utilize prompt-based interaction with LLMs to perform various tasks. These methods can be broadly categorized into In-Context Learning (ICL), Chain-of-Thought (CoT) reasoning, and Task Instruction Prompting. In practice, these techniques are often used in combination to complement one another. For simplicity, we classify such hybrid methods based on the primary technique emphasized in each study. These approaches commonly utilize diverse data sources, including log data, incident reports, and metrics data.

\textbf{In-Context Learning (ICL).}  In-Context Learning involves providing LLMs with examples or context within the prompt to guide the model in task execution. This approach allows the model to infer patterns from the examples and produce outputs aligned with desired results, making it particularly useful for tasks requiring format adherence or specific outputs.

Many studies utilize ICL in log parsing tasks. Works such as LILAC~\cite{jiang2024lilac}, LLMParser~\cite{ma2024llmparser}, Lemur~\cite{guo2024lemur}, DivLog~\cite{xu2024divlog}, LogParser-LLM~\cite{zhong2024logparser}, ECLIPSE~\cite{zhang2024eclipse}, Free~\cite{xiao2024free}, LogBatcher~\cite{xiao2024stronger}, and LogPPT~\cite{le2023log} effectively pair ICL with cache mechanisms to improve parsing accuracy and efficiency. For log-based anomaly detection, LogPrompt~\cite{liu2024interpretable}, LogEval~\cite{cui2024logeval}, and Semirald~\cite{sun2024semirald} adopt ICL to guide the model in identifying anomalous patterns within structured or semi-structured data.

In the context of incident reports, ICL has been widely used to address tasks like root cause analysis. For instance, RCACopilot~\cite{chen2024automatic} and Xpert~\cite{jiang2024xpert} leverage ICL to structure diagnostic processes, ensuring accurate identification of causative factors. For metrics-based methods, LasRCA~\cite{han2024potential} combines one-shot learning with smaller models as classifiers to enhance metrics-based anomaly detection. Despite its focus on metrics, the method integrates incident reports for failure identification, showcasing the adaptability of ICL in leveraging both structured and unstructured data across various AIOps scenarios. Similarly, AnomalyLLM~\cite{liu2024anomalyllm} employs a previously fine-tuned LLaMA2-7B model and applies ICL to effectively detect metrics-based anomalies, further demonstrating the versatility of this approach.

\textbf{Chain of Thoughts (CoT).}  Chain of Thought prompting enhances LLMs' capabilities in multi-step reasoning tasks by structuring prompts to guide the model through intermediate steps before arriving at the final output. This method is particularly advantageous for tasks requiring logical decomposition or multi-step processing.

For time-series forecasting, LSTPrompt~\cite{liu2024lstprompt} decomposes predictions into short-term and long-term sub-tasks, utilizing CoT techniques to generate tailored prompts for each step. Similarly, LogGPT~\cite{liu2024interpretable} leverages CoT prompting to refine anomaly detection in logs, enabling the model to better capture nuanced semantic features.

In causal discovery and root cause analysis, RealTCD~\cite{li2024llm} applies CoT prompting to identify causative relationships between events. LM-PACE~\cite{zhang2024lm} uses CoT techniques to enhance GPT-4’s diagnostic capabilities in analyzing incident reports, while~\cite{hamadanian2023holistic} extends this approach to generating mitigation solutions from incident reports. These examples highlight the effectiveness of CoT prompting in addressing complex analytical tasks within AIOps.

\textbf{Task Instruction Prompting.} Task Instruction Prompting involves providing explicit, detailed instructions to LLMs to guide task execution. This method is particularly effective for zero-shot or simple tasks where direct instructions suffice to produce satisfactory results.

Early works in AIOps often relied on this approach for anomaly detection tasks. For instance, LogGPT~\cite{qi2023loggpt} and Priyanka et al.~\cite{quan2023heterogeneous} conduct studies on zero-shot log-based anomaly detection using ChatGPT, revealing its limited performance without further fine-tuning or specialized techniques. Metrics-based anomaly detection has also been tackled through instruction-based approaches. Sigllm\cite{othman2023fostering} and TabLLM~\cite{ahmed2023recommending} integrate time-series-to-text conversion modules and explicitly prompt LLMs to identify anomalies within the data.

In root cause analysis and assisted remediation tasks, Achraf et al.~\cite{othman2023fostering} demonstrate the use of zero-shot prompt-based script generation to support solution development, while Toufique et al.~\cite{ahmed2023recommending} compared zero-shot prompting with parameter-efficient tuning for root cause analysis and solution recommendation based on incident reports. Andres et al.~\cite{quan2023heterogeneous} employe GPT-3.5 in a zero-shot setup for log-based root cause analysis, underscoring the flexibility and adaptability of task instruction prompting in diverse applications.

\subsection{Knowledge-based Approach}

In addition to directly utilizing prompts for AIOps tasks, knowledge-based approaches enhance the performance of LLMs by integrating historical data or external knowledge sources. These approaches can be broadly categorized into two main types: Retrieval-Augmented Generation (RAG) and Tool-Augmented Generation (TAG). Methods based on historical data typically leverage log data, metrics data, and incident reports. In contrast, methods that rely on external knowledge sources often incorporate trace data, system information, and source code to enrich the model’s reasoning capabilities.

\textbf{Retrieval Augmented Generation (RAG).} Retrieval Augmented Generation enhances the model's outputs by retrieving relevant information from external knowledge bases or databases. This approach enables LLMs to incorporate up-to-date or domain-specific knowledge that is not part of their pre-training. In AIOps, RAG is widely applied to access historical logs, incident reports, or technical documentation, helping models provide contextually informed and reliable solutions.

Some works employ RAG specifically to retrieve historical data, using it as a reference for current AIOps tasks. These methods can be considered an improvement on the aforementioned ICL-based approaches, leveraging specialized similarity retrieval techniques to select the most relevant "shots" for the large language model. Xuchao et al.~\cite{zhang2024automated} developed an incident report summarization system that converts report content into a 768-dimensional dense vector space. In production environments, similar vectors are retrieved based on proximity to the current incident and passed to GPT-4 for root cause analysis. LM-PACE~\cite{zhang2024lm} uses similarity retrieval to identify relevant historical incidents, jointly retrieving corresponding recommended root causes to assist in confidence calibration. Xpert~\cite{jiang2024xpert} retrieves similar incident reports to recommend repair commands. Oasis~\cite{jin2023assess} generates summaries of current incidents by retrieving historical patterns between incidents. Methods such as RAGLog~\cite{pan2023raglog} and LogRAG~\cite{zhang2024lograg} use historical log retrieval to enhance log-based anomaly detection. MonitorAssistant~\cite{yu2024monitorassistant} retrieves similar historical monitor metrics and anomaly labels to support anomaly detection.

Other works utilize RAG to retrieve external knowledge, providing richer data and domain expertise to strengthen AIOps solutions. Drishti et al.~\cite{goel2024x} retrieve information across different stages of the software development lifecycle (SDLC) to enhance the LLM’s root cause analysis capabilities. LLMDB~\cite{zhou2024llm} and DB-GPT~\cite{xue2023db} integrate database-specific knowledge (e.g., developer documentation and user manuals) to optimize anomaly detection and diagnosis in database systems. LogExpert~\cite{wang2024logexpert} retrieves domain features similar to anomalous logs to aid in generating mitigation solutions. RAG4ITOps~\cite{zhang2024rag4itops} retrieves domain-specific knowledge bases from cloud computing to support IT operations knowledge Q\&A.

\textbf{Tool Augmented Generation (TAG).} Tool Augmented Generation equips large language models with the ability to interact with external tools, APIs, or software systems to assist in completing tasks. In AIOps, TAG-based methods enable LLMs to query monitoring systems, execute diagnostic scripts, or use debugging tools, generating more accurate and actionable results. This approach is particularly effective for automating complex workflows that demand precise execution and real-time data interaction.

In AIOps, some TAG-based methods use tools to aid in failure identification and classification. RCAgent~\cite{wang2023rcagent} utilizes information-gathering and analytical tools to perform root cause analysis for Apache Flink. RCACopilot~\cite{chen2024automatic} focuses exclusively on information-gathering tools to assist in incident report-based root cause analysis.

Other works employ tools to execute specific operations, which are primarily applied in assisted remediation tasks. Charles et al.~\cite{cao4741492managing} introduce a GPT-based AI agent capable of competently using 150 distinct tools across nine categories, from file manipulation to programming compilation, to manage Linux servers. NetLM~\cite{wang2023network} leverages various traffic control tools to automate network management. Pitikorn et al.~\cite{khlaisamniang2023generative} employ code execution tools to automatically execute Python scripts, enabling autonomous cloud system repairs. ChatOps4Msa~\cite{wang2023low}, Pesl et al.~\cite{pesl2023uncovering}, and Marco et al.~\cite{aiello2023service} utilize DevOps tools to explore automated service composition.

A subset of works integrates both analytical and execution tools, blending their capabilities. Honghao et al.~\cite{shi2024enhancing} propose a method that combines analytical and execution tools to perform early warning, troubleshooting, and repair tasks for AI clusters. LLexus~\cite{las2024llexus} integrates script tools and semantic tools based on troubleshooting guides (TSGs) to mitigate failures in cloud systems. AIOpsLab~\cite{shetty2024building} leverages traces, metrics, and log data in combination with various information-gathering and automated execution tools to localize and resolve failures in microservice clusters.

\section{RQ4: Evaluating LLM-based AIOps}
\label{sec:rq4}

In this section, we discuss the emergence of new evaluation methodologies in the era of large language models for AIOps. These advancements are reflected in two key areas: the development of new evaluation metrics and the creation of novel datasets.

\subsection{Emerging Evaluation Metrics}

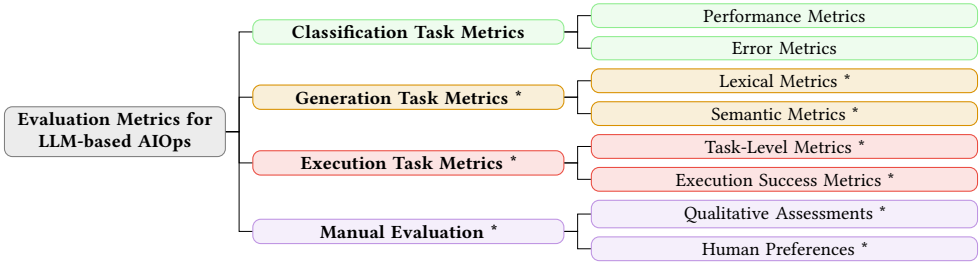
\begin{figure}[htbp]
	\centering
	\tikzset{
		my node/.style={
			draw,
			align=center,
			thin,
			text width=1.2cm, 
			rounded corners=3,
		},
		my leaf/.style={
			draw,
			align=center,
			thin,
			text width=8.5cm, 
			rounded corners=3,
		}
	}
	\forestset{
		every leaf node/.style={
			if n children=0{#1}{}
		},
		every tree node/.style={
			if n children=0{minimum width=1em}{#1}
		},
	}
	\begin{forest}
		nonleaf/.style={font=\bfseries\scriptsize},
		for tree={%
			every leaf node={my leaf, font=\scriptsize},
			every tree node={my node, font=\scriptsize, l sep-=4.5pt, l-=1.pt},
			anchor=west,
			inner sep=2pt,
			l sep=10pt, 
			s sep=3pt, 
			fit=tight,
			grow'=east,
			edge={ultra thin},
			parent anchor=east,
			child anchor=west,
			if n children=0{}{nonleaf}, 
			edge path={
				\noexpand\path [draw, \forestoption{edge}] (!u.parent anchor) -- +(5pt,0) |- (.child anchor)\forestoption{edge label};
			},
			if={isodd(n_children())}{
				for children={
					if={equal(n,(n_children("!u")+1)/2)}{calign with current}{}
				}
			}{}
		}
		[Evaluation Metrics for \\ LLM-based AIOps, draw=gray, fill=gray!15, text width=2.8cm, text=black
		[Classification Task Metrics, color=lightgreen, fill=lightgreen!15, text width=4cm, text=black
		[Performance Metrics, color=lightgreen, fill=lightgreen!15, text width=5cm, text=black]
		[Error Metrics, color=lightgreen, fill=lightgreen!15, text width=5cm, text=black]
		]
		[Generation Task Metrics *, color=harvestgold, fill=harvestgold!15, text width=4cm, text=black
		[Lexical Metrics *, color=harvestgold, fill=harvestgold!15, text width=5cm, text=black]
		[Semantic Metrics *, color=harvestgold, fill=harvestgold!15, text width=5cm, text=black]
		]
		[Execution Task Metrics *, color=carminepink, fill=carminepink!15, text width=4cm, text=black
		[Task-Level Metrics *, color=carminepink, fill=carminepink!15, text width=5cm, text=black]
		[Execution Success Metrics *, color=carminepink, fill=carminepink!15, text width=5cm, text=black]
		]
		[Manual Evaluation *, color=brightlavender, fill=brightlavender!15, text width=4cm, text=black
		[Qualitative Assessments *, color=brightlavender, fill=brightlavender!15, text width=5cm, text=black]
		[Human Preferences *, color=brightlavender, fill=brightlavender!15, text width=5cm, text=black]
		]
		]
	\end{forest}
	\caption{Evaluation Metrics for LLM-based AIOps (Metrics marked with * are new metrics that have emerged in the era of LLMs)}
	\label{fig: metrics}
\end{figure}

With the evolution of AIOps tasks in the LLM era, the range of evaluation metrics has expanded significantly. As shown in Figure~\ref{fig: metrics}, current evaluation metrics can be categorized into four types: classification task metrics, generation task metrics, execution task metrics, and manual evaluation. Notably, the latter three have emerged as new additions in the LLM era.

\subsubsection{Classification Task Metrics} This category of metrics originates from traditional AIOps tasks and continues to be utilized in some works in the era of large language models (LLMs)~\cite{sun2023test, gruver2024large, cao2023tempo, goel2024x, pan2023raglog, jin2023time, chang2023llm4ts, liao2024timegpt, rasul2023lag, dasdecoder, liu2024lstprompt, jiang2024lilac, guo2024lemur, le2023log, xu2024divlog}. These metrics can be broadly classified into two types: \textbf{Performance Metrics} and \textbf{Error Metrics}.

\textbf{Performance Metrics.} These metrics evaluate the accuracy of model classifications, including precision, recall, F1-score, accuracy, and AUC-ROC. They are frequently used for tasks such as log-based anomaly detection and failure category classification. In failure localization tasks, metrics like Acc@ (accuracy of the top-N predictions) are also employed. For log parsing, additional metrics such as grouping accuracy (GA) are utilized to measure performance.

\textbf{Error Metrics.} These metrics are primarily used for regression tasks, quantifying the difference between predicted and actual values. Examples include MAE (Mean Absolute Error), MSE (Mean Squared Error), RMSE (Root Mean Squared Error), and MAPE (Mean Absolute Percentage Error). Such metrics are commonly applied in evaluating large time-series models for tasks like failure prediction and anomaly detection.

\subsubsection{Generation Task Metrics} This category of metrics was not frequently used in traditional AIOps. However, with the integration of large language models, a variety of generative tasks have emerged, such as root cause report generation and mitigation solution generation. As a result, generation-specific metrics are now essential for evaluating these methods~\cite{wang2023rcagent, zhang2024automated, shi2023shellgpt, ahmed2023recommending, jiang2024xpert, meng2023logsummary, liu2024loglm, wang2024logexpert, roy2024exploring, mudgal2024crasheventllm, namrud2024kubeplaybook}. These metrics can be classified into two groups: \textbf{Lexical Metrics} and \textbf{Semantic Metrics}.

\textbf{Lexical Metrics.} These metrics focus on surface-level textual similarity, comparing the generated output with reference texts based on word overlap. Examples include BLEU (Bilingual Evaluation Understudy), ROUGE (Recall-Oriented Understudy for Gisting Evaluation), and METEOR (Metric for Evaluation of Translation with Explicit ORdering). These metrics are particularly useful for tasks where adherence to specific language or terminology is critical, such as generating standardized incident reports or summaries.

\textbf{Semantic Metrics.} These metrics assess the meaning and contextual relevance of the generated text rather than its exact lexical similarity. Examples include BERTScore, BLEURT (Bilingual Evaluation Understudy with Representations from Transformers) and NUBIA (NeUral Based Interchangeability Assessor). These metrics are crucial for evaluating tasks where the primary concern is the meaningfulness or contextual appropriateness of the output, such as mitigation suggestions or explanations for system failures.

\subsubsection{Execution Task Metrics} This category of metrics has emerged primarily with the rise of large language models, as they are designed to evaluate assisted remediation tasks, particularly the effectiveness of script generation and automatic execution~\cite{sarda2023leveraging, cao4741492managing, sarda2023adarma, minna2024analyzing, sarda2024leveraging}. These metrics can be broadly divided into two groups: Task-Level Metrics and Execution Success Metrics.

\textbf{Task-Level Metrics.} These metrics assess the quality and correctness of individual tasks or blocks within a generated script. They are particularly important for evaluating script generation tasks, where the output consists of multiple functional components that need to be validated independently. Examples include:

\begin{itemize}
	\item \textbf{Functional Correctness (FC)}: Measures whether each generated task or script successfully fulfills its intended functionality. This often involves running the tasks and verifying the results against ground truth or predefined benchmarks.
	\item \textbf{Average Correctness (AC)}: Evaluates the proportion of successfully executed tasks within a generated script. For example, in systems like Ansible, tasks represent units of work such as installing a package, configuring services, or copying files. The AC metric calculates the average success rate of all tasks, reflecting the overall correctness of the script.
\end{itemize}

\textbf{Execution Success Metrics.} These metrics evaluate the success of executing the generated script or code in its entirety, ensuring the intended outcome is achieved. They focus on operational effectiveness, particularly for automation tasks. Examples include:

\begin{itemize}
	\item \textbf{Execution Success Rate}: Measures the percentage of generated scripts or workflows that execute without errors and achieve the desired outcome.
	\item \textbf{Correct Refactorings}: Assesses whether modifications or refinements to an existing script or workflow (e.g., bug fixes or performance optimizations) lead to successful and improved executions.
\end{itemize}

\subsubsection{Manual Evaluation} While the previously discussed metrics are mathematically rigorous and provide objective assessments, they are often insufficient to fully evaluate LLM-based AIOps methods. The powerful generative capabilities of large language models frequently require human involvement to assess their outputs~\cite{zhang2024lm, mudgal2023assessment, othman2023fostering, shan2024face, roy2024exploring, yu2024monitorassistant, las2024llexus, shetty2024building}. This is especially true for tasks such as root cause report generation and mitigation solution generation, where the outputs often lack fixed labels or have inherently flexible ground truth. Manual evaluation methods can be broadly categorized into two groups: \textbf{Qualitative Assessments} and \textbf{Human Preferences}.

\textbf{Qualitative Assessments.} .These assessments involve human experts objectively evaluating the results against predefined qualitative grading criteria. The goal is to identify specific shortcomings and categorize the outputs based on quality or error types. For exmaple, a root cause report can be evaluted by the following levels~\cite{roy2024exploring}:

\begin{itemize}
	\item Precise: Results are accurate, relevant, and fully aligned with the task requirements.
	\item Imprecise: Results are partially correct but contain minor inaccuracies or missing details.
	\item Hallucination: Results include fabricated or irrelevant information not grounded in the input data.
	\item Reasoning Error: Logical errors or invalid inferences made in the generated output.
	\item Retrieval Error: Incorrect or incomplete use of retrieved information during generation.
\end{itemize}

\textbf{Human Preferences.} This type of manual evaluation focuses on subjective assessments of the utility, relevance, and overall usefulness of the generated results. Unlike strict pre-defined criteria, evaluators provide judgments based on personal perspectives or case-specific requirements. A classic example is case studies, where human evaluators assess outputs in real-world scenarios to determine whether the generated solutions are practically useful and actionable.

\subsection{Development of Novel Datasets}

In the era of large language models, several novel datasets have been specifically introduced for AIOps. It is important to note that this section focuses exclusively on works that explicitly design datasets to cater to large language models. In other words, we exclude datasets proposed in recent years that, although utilized in LLM-based works, were not originally developed with a specific focus on LLMs. Notably, most of these datasets are concentrated on assisted remediation tasks due to the abundance of new challenges in this area.

Only a few works have introduced datasets targeting failure prediction and root cause analysis tasks. For example, LogEval~\cite{cui2024logeval} was proposed to evaluate LLM capabilities across various log analysis tasks, including log parsing, log anomaly detection, log fault diagnosis, and log summarization. It assesses these tasks using 4,000 publicly available log data entries and employs 15 different prompts for each task.

In the assisted question task, OpsEval~\cite{liu2023opseval} is introduced as a comprehensive task-oriented benchmark designed specifically for LLMs. This dataset includes 7,184 multiple-choice questions and 1,736 question-answering (QA) formats in both English and Chinese. It represents the first effort to evaluate LLMs' proficiency across various critical scenarios and skill levels. Additionally, OpsEval provides an online leaderboard and a continuously updated dataset, along with the leaderboard itself~\footnote{https://opseval.cstcloud.cn/content/leaderboard}. Similarly, OWL-bench~\cite{guo2023owl} has established a dataset for the operation and maintenance (O\&M) domain. This dataset spans nine O\&M-related subdomains, showcasing the diversity and hierarchical structure of LLM capabilities within the O\&M field.

In the script generation task, KubePlaybook~\cite{namrud2024kubeplaybook} introduced a curated dataset comprising 130 natural language prompts for generating automation-focused remediation code scripts.

Additionally, some works have proposed benchmark frameworks encompassing the entire AIOps lifecycle. For instance, AIOpsLab~\cite{chenaiopslab}, developed by Microsoft, provides a prototype implementation leveraging an agent-cloud interface. This framework orchestrates applications, injects real-time failures using chaos engineering, and interfaces with agents to localize and resolve failures.

\section{Challenges and Future Directions}
\label{sec:challenges}

While large language models have significantly enhanced AIOps, numerous challenges still remain to be addressed. Many of these challenges stem from inherent limitations of LLMs themselves, such as hallucinations, inconsistent outputs, context limitations, and others. These general issues have been widely studied in the field of natural language processing (NLP) and fall outside the scope of this paper.

However, beyond these general challenges, LLM-based AIOps presents a set of unique and domain-specific issues that require further investigation. These include the time-efficiency and cost-effectiveness of LLM-based AIOps solutions, the in-depth utilization of diverse failure data sources, the generalizability and adaptability of models during software evolution, and the integration with existing AIOps toolchains. These domain-specific challenges will be discussed in detail in the following sections.

\subsection{Time-Efficiency and Cost-effectiveness of LLM-based AIOps Solutions}

Large language models require substantial computational resources and energy, leading to high operational costs. The training and deployment of LLMs involve extensive use of GPUs or TPUs, making them less accessible for many organizations. Additionally, the inference costs associated with LLMs can be prohibitive, especially when real-time or near-real-time responses are required. This poses significant hurdles, particularly for small to medium-sized enterprises or in scenarios where computational resources are constrained.

Among the tasks of AIOps, the computational overhead of LLMs is most critical for failure perception tasks. These steps theoretically need to run continuously during the operation of a software system and require high real-time performance. For example, if the anomaly detection time window is set to 10 seconds, meaning an anomaly detection process is triggered every 10 seconds, the anomaly detection model must infer results within 1 second to promptly notify OCEs in case of anomalies. Some non-LLM-based approaches have focused on addressing this issue~\cite{jia2021logflash, lin2024fastlogad}, but currently, no LLM-based work has adequately tackled this problem. However, this issue is becoming increasingly significant in the LLM era.

On the other hand, while the higher computational efficiency and cost associated with LLMs may be acceptable for root cause analysis and auto remediation (since these steps do not need to run continuously within the software system), this issue still requires careful consideration. Excessive computational costs could potentially make LLM-based approaches less effective than using small-scale models in combination with OCEs. For example, a well-designed smaller model augmented by human expertise might achieve comparable results at a fraction of the cost and resource consumption, making it a more practical and scalable solution in certain contexts.

In summary, balancing computational cost and performance is crucial for the application of LLMs in AIOps. If the cost of running LLMs outweighs their benefits, organizations may find it more advantageous to leverage traditional models or hybrid approaches that combine the strengths of smaller-scale models and human oversight. Therefore, future research should focus on optimizing the computational efficiency of LLMs for AIOps, possibly by exploring methods that organically combine small-scale models, large language models, and OCEs. This integrated approach could provide a time-efficient and cost-effective solution while maximizing the strengths of each component.

\subsection{In-depth Usage of More Diverse Failure Data Sources}

Several important data sources have yet to be utilized in LLM-based approaches for AIOps. As illustrated in Section~\ref{sec:rq1}, there are three crucial types of system-generated data in software system runtime: metrics, logs, and traces. While there has been significant progress in utilizing metrics and logs, no current work has effectively incorporated traces data.

Traces data is particularly valuable as it provides insights into the calls between nodes in a software system cluster and the interactions within internal components. The lack of LLM-based approaches using traces data can be attributed to its high complexity and volume, making it challenging to integrate trace data with LLMs. Traces often involve detailed and nested sequences of events that need to be represented in a manner that LLMs can effectively understand and utilize. However, traces data remains a crucial resource for comprehensive system monitoring and AIOps. Future research should focus on developing methods to effectively integrate traces data into LLM-based AIOps methods.

On the other hand, while many failure perception works have been based on logs, a significant portion of these works utilize pre-trained models like T5 and GPT-2, which are not particularly large in scale~\cite{xue2023promptcast, huang2024demystifying, sun2024art, le2024prelog}. These approach result in relatively high training costs. Only a small fraction of works have utilized larger models such as GPT-3.5, but these studies are primarily based on simple datasets~\cite{egersdoerfer2023early, pan2023raglog, qi2023loggpt}. For these datasets, failure perception can often be achieved with traditional machine learning models~\cite{landauer2023critical, zhang2024reducing}, which diminishes the perceived superiority of LLM-based methods. Therefore, future research should explore more effective ways to apply logs to LLM-based approaches. One feasible solution could be prompt embedding, which has been widely used in metrics-based work.

Lastly, in LLM-based root cause analysis, many works have used incident reports as the primary data source and achieved excellent results~\cite{zhang2024automated, goel2024x, zhang2024lm, khlaisamniang2023generative, ahmed2023recommending, jiang2024xpert, kuang2024knowledge, hamadanian2023holistic, jin2023assess, roy2024exploring}. However, this approach disrupts the automation flow of AIOps, as root cause analysis should be triggered by failure perception. Therefore, future work should focus on using system-generated data to trigger failure perception, generate corresponding incident reports, and then apply root cause analysis methods to analyze these generated incident reports.

\subsection{Generalizability and Model Adaptability in Software Evolution}

A key challenge in traditional ML- and DL-based approaches to AIOps lies in significant performance degradation when models are applied to different software systems or encounter modifications to the original system. Techniques such as meta-learning and online learning have sought to mitigate this by enabling models to retrain efficiently with minimal new data, thus restoring their effectiveness in evolving environments.

In contrast, LLM-based approaches, leveraging the inferential power of large language models, are expected to exhibit high generalizability and adaptability due to extensive pre-training on diverse textual data. However, empirical validation of this expectation remains limited. While some foundation models demonstrate cross-platform effectiveness, many works, particularly those using prompt-based methods, lack systematic evaluation of generalizability and adaptability in real-world software evolution scenarios.

To address this gap, future research must emphasize rigorous testing of LLMs' ability to generalize across diverse systems and maintain or recover performance amid software changes. This involves designing systematic experiments to simulate varying software environments and assessing model resilience through incremental fine-tuning or continuous learning paradigms. Furthermore, exploring the intersection of advanced LLM architectures, transfer learning techniques, and adaptive learning algorithms will be pivotal in ensuring robust, scalable, and adaptable solutions for AIOps.

\subsection{Integration with Existing AIOps Toolchains}

Current LLM-based AIOps solutions predominantly focus on constructing entirely new methods, often disregarding the value of existing smaller model-based approaches. However, considering the diverse tasks within AIOps, such as failure prediction, root cause analysis, and assisted remediation, a more effective approach would involve integrating LLMs with existing toolchains to enhance overall performance.

In failure prediction, smaller models or rule-based systems could quickly process high-frequency, structured data like logs or metrics, providing initial anomaly detection. LLMs, with their contextual reasoning and flexibility, could then interpret and correlate these anomalies with broader system behaviors, offering deeper insights. Similarly, in root cause analysis, traditional log parsers or anomaly detectors might identify potential fault patterns, which LLMs could augment by generating detailed reports or explanations that draw on diverse data sources and historical knowledge.

In assisted remediation, existing automation frameworks (e.g., Ansible or Kubernetes operators) could handle script execution and simple task orchestration, while LLMs contribute by generating, validating, or refining automation scripts. This division of labor ensures efficiency, as routine or repetitive tasks are offloaded to smaller models or tools, while LLMs focus on tasks requiring advanced reasoning or generation.

Future research should focus on developing interoperable frameworks that seamlessly integrate LLMs with smaller models, legacy systems, and specialized tools. Such modular integration not only leverages the strengths of traditional methods but also reduces the computational burden and operational risks of relying solely on large models—ensuring scalable, cost-efficient, and easily adoptable solutions across diverse AIOps scenarios.

\section{Conclusion}

LLMs are bringing significant changes to the field of AIOps. Their powerful generative capabilities fundamentally reshape many AIOps tasks and methodologies. In this survey, we analyzed the emerging utilization of LLMs for AIOps, focusing on research related to LLMs (models with more than 1 billion parameters). We examined the diverse failure data sources applied in this field, including advanced LLM-based processing techniques for legacy data and the utilization of new data sources enabled by LLMs (RQ1). We also investigated the evolving tasks in AIOps with LLMs, emphasizing the emergence of new tasks and analyzing the distribution of publications across different tasks (RQ2). Following this, we explored the various LLM-based methods applied to these tasks (RQ3). Lastly, we reviewed the evaluation methodologies in AIOps adapted to the integration of LLMs (RQ4).

Despite the progress made, several areas require further exploration. These include the time-efficiency and cost-effectiveness of LLM-based AIOps solutions, deeper utilization of diverse data sources, generalizability and model adaptability during software evolution, integration with existing AIOps toolchains.

In conclusion, the realm of AIOps in the era of LLMs is a dynamically advancing field teeming with both promise and challenges. Its progression is vital to underpinning the stability and reliability of software systems. By addressing the challenges outlined in this survey, LLM-powered methodologies for AIOps are poised for broader, more impactful real-world implementations. This advancement will not only enhance the resilience and efficiency of modern software ecosystems but also pave the way for intelligent, adaptive, and future-ready solutions to manage software failures effectively.

\begin{acks}
 This work was supported by the National Key R\&D Research Fund of
 China (2021YFF0704202).
\end{acks}

\bibliographystyle{ACM-Reference-Format}
\bibliography{sample-base}

\end{document}